\documentclass[10pt]{iopart}
\makeatletter
\@namedef{ver@amsmath.sty}{}
\makeatother
\usepackage{amstext}
\usepackage{graphics}% Include figure files
\usepackage{graphicx}% Include figure files
\usepackage{cite}
\usepackage{epsfig}
\usepackage{epstopdf}
\usepackage{subfigure}
\usepackage{braket}
\DeclareMathAlphabet{\mathpzc}{OT1}{pzc}{m}{it}
\usepackage[dvips]{color}
\usepackage{gensymb}
\usepackage{dcolumn}% Align table columns on decimal point
\usepackage{bm}% bold math
\usepackage{amssymb}
\usepackage{iopams}
\usepackage{amsmath}
\usepackage{indentfirst}
\usepackage{booktabs}
\usepackage{multirow}
\usepackage{colortbl}
\usepackage{url}
\usepackage{caption}
\begin{document}

\title{Large bulk photovoltaic effect and second-harmonic generation in few-layer 
pentagonal semiconductors PdS$_2$ and PdSe$_2$}
\author{Vijay Kumar Gudelli$^1$ and Guang-Yu Guo$^{2,1}$}
\address{$^1$ Physics Division, National Center for Theoretical Sciences, Taipei 10617, Taiwan}
\address{$^2$ Department of Physics and Center for Theoretical Physics, National Taiwan University, Taipei 10617, Taiwan}
\ead{gyguo@phys.ntu.edu.tw}

\vspace{10pt}
\begin{indented}
%\date{\today}
\item[\today]
\end{indented}

\begin{abstract}
Recently, atomically thin PdSe$_2$ semiconductors with rare pentagonal Se-Pd-Se monolayers
were synthesized and were also found to possess superior properties such as ultrahigh air stability,
tunable band gap and high carrier mobility, thus offering a new family of two-dimensional (2D)
materials for exploration of 2D semiconductor physics and for applications in
advanced opto-electronic and nonlinear photonic devices.  
In this work, we systematically study the nonlinear optical (NLO) responses 
[namely, bulk photovoltaic effect (BPVE), second-harmonic generation (SHG) and linear electric-optic (LEO) effect] 
of noncentrosymmetric bilayer (BL) and four-layer (FL) PdS$_2$ and PdSe$_2$ 
by applying the first-principles density functional theory with the generalized 
gradient approximation plus scissors-correction.
First of all, we find that these few-layer PdX$_2$ (X = S and Se) exhibit prominent BPVE.
In particular, the calculated shift current conductivity is in the order of 130 $\mu$A/V$^2$, 
being very high compared to known BPVE materials. Similarly, 
their injection current susceptibilities are in the order of 100$\times$10$^8$A/V$^2$s, again being large.
Secondly, the calculated SHG coefficients ($\chi^{(2)}$) of these materials are also large,
being one order higher than that of the best-known few-layer group 6B transition metal dichalcogenides.
For example, the maximum magnitude of $\chi^{(2)}$ can reach 1.4$\times$10$^3$ pm/V 
for BL PdSe$_2$ at 1.9 eV and 1.2$\times$10$^3$ pm/V at 3.1 eV for BL PdS$_2$. 
Thirdly we find significant LEO coefficients for these structures in the low photon energy.
All these indicate that 2D PdX$_2$ semiconductors will find promising NLO applications in light signal modulators, 
frequency converters, electro-optical switches and photovoltaic solar cells.
Fourthly, we find that the large BPVE and SHG of the few-layer PdX$_2$ structures 
are due to strong intralayer directional covalent bonding and also 2D quantum confinement.
%their low crystal symmetry, strong intralayer directional covalent bonding
%Finally, the strong NLO responses of BL and FL structures of PdX$_2$ are attributed to
%strong intralayer directional covalent bonding and also 2D quantum confinement.
%and relatively strong interlayer interaction.  
Finally, we also discuss the prominent features of these NLO spectra of these materials 
in terms of their electronic structure and optical dielectric functions.
\end{abstract}


\section{Introduction}
Because of their extraordinary electronic and optical properties, two-dimensional (2D) materials such as graphene, 
atomically thin (few-layer) transition metal dichalcogenides (TMDCs) and black phosphorus, have attracted
an enormous amount of interest in recent years, finding diverse applications in electronic, opto-electronic
and nonlinear photonic devices with superior performances. 
Among them, group 6B TMDC semiconductors with chemical formula MX$_2$ (M = Mo, W; X = S, Se) and each
layer made up of a 2D hexagonal array of M atoms sandwiched between the similar arrays of X atoms, constitute
a particularly interesting family of 2D materials. In particular, they were found to exhibit an indirect to direct
band gap transition when they were thinned down to a monolayer (ML)~\cite{Mak2010}. 
This makes the MX$_2$ MLs semiconductors with 
a direct band gap, thus becoming promising materials for, e.g., electro-optical devices with efficient light
emission~\cite{Mak2010} and field effect transistors with high on-off ratios~\cite{Radisavljevic2011}. 
Furthermore, these hexagonal 2D MX$_2$ materials
with an odd layer-number lack the spatial inversion symmetry, 
although their bulk crystals are centrosymmetric.
This broken inversion symmetry makes them
exhibit novel properties of 
fundamental and technological interests, especially second-order nonlinear optical (NLO) responses such as
second-harmonic generation (SHG)~\cite{Shen2003,Boyd2003,Wang2015b} and bulk photovoltaic effect (BPVE)~\cite{Sturman1992}. 

Stimulated by the recent fabrications of few-layer PdSe$_2$ via molecular beam epitaxy, 
chemical vapor deposition and mechanical 
exfoliation~\cite{Oyedele2017,Chow2017,Li2018}, palladium-based TMDC 2D materials 
have also attracted much attention in the past five years~\cite{Pi2019}.
Indeed, these few-layer Pd-based TMDC materials were found to exhibit some desired 
properties for applications, such as tunable band gap, high carrier mobility, anisotropy,
enhanced thermoelectric property and ultrahigh air 
stability~\cite{Oyedele2017,Lan2019,Pi2019,Liang2019}. 
There are several distinct differences between group 6B TMDC and Pd-based TMDC 2D materials. 
In particular, compared with Mo and W atoms, Pd atoms have a nearly filled $d$-shell
and thus there are stronger hybridization between Pd $d$-orbitals and chalcogen $p$-orbitals in Pd-based TMDC 2D materials. 
This result in stronger covalent bonding within each PdX$_2$ (X = S and Se) layer as well as stronger interlayer binding.  
The latter gives rise to layer-number dependent properties especially band gap size~\cite{Oyedele2017,Pi2019}.
In contrast to groups 5B and 6B TMDCs, bulk PdS$_2$ and PdSe$_2$ crystallize in the 
orthorhombic layered structure [see Fig. 1(a)] with the centrosymmetric $Pbca$ space group~\cite{Gronvold1957,Soulard2004}.
Furthermore, in each PdX$_2$ layer, a Pd atom bonds with four chalcogen atoms, and 
Pd and chalcogen atoms form a buckled pentagonal layered structure [see Fig. 1(c)]~\cite{Oyedele2017,Liang2019}. 
Depending on their thickness (i.e., layer-number), few-layer PdX$_2$ structures 
have different symmetries compared with their bulk crystals. 
Interestingly, 2D PdX$_2$ materials with an even layer-number crystallizes  
in a noncentrosymmetric structure with space group $Pca2_1$ and point group symmetry of $C_{2v}$ (or $mm2$),
while 2D PdX$_2$ materials with an odd layer-number form a centrosymmetric structure of
space group $P2_1/c$~\cite{Oyedele2017,Pi2019,Kuklin2019}. 
Note that 2D group 6B TMDC structures with an odd-layer number
are noncentrosymmetric, while that with an even layer-number have the inversion symmetry.  
Therefore, as for few-layer group 6B TMDC materials with an odd layer-number, 
2D PdX$_2$ structures with an even layer-number are expected to show
second-order NLO properties. 
Indeed, strong angle-dependent SHG signals in bilayer (BL) 
and four-layer (FL) PdSe$_2$ were observed recently~\cite{Puretzky2018}, although
the precise SHG susceptibility was not determined. 
Motivated by these exciting developments,
in this work we perform a systematic theoretical study of the NLO responses
of BL and FL PdX$_2$, based on first-principles density functional theory (DFT) calculations.

\begin{figure}[htb]
\begin{center}
\includegraphics[width=12cm]{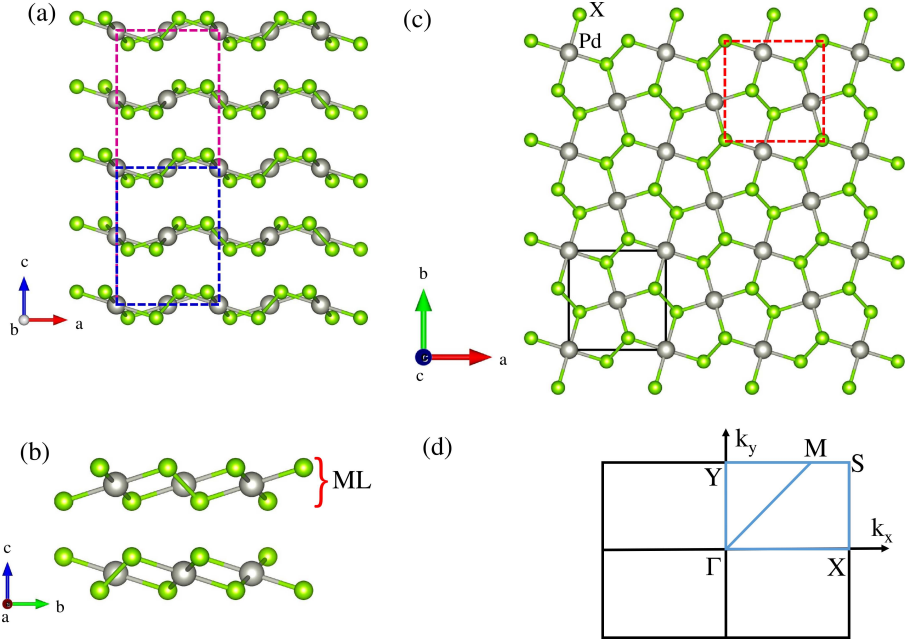}
\end{center}
\caption{(a) Crystal structure of bulk PdX$_2$ (side view along the $b$-axis), where blue dashed lines 
(magenta dashed lines) represent the unit cell of BL PdX$_2$ (FL PdX$_2$).
%and magenta-dashed line represents the unit cell of FL PdX$_2$ 
(b) Crystal structure of BL PdX$_2$ (side view along the $a$-axis).
(c) Top view of the PdX$_2$ structures. Here black solid lines (red dashed lines) show the unit cell 
(the pentagon) of a PdX$_2$ layer. (d) 2D Brillouin zone (BZ) for BL and FL PdX$_2$ structures.
The blue rectangle denotes the irreducible BZ wedge. The conduction band minimum
is located at a general $k$-point on the $\Gamma$-M line (see the text). }
\label{structure}
\end{figure}

In this paper, we focus on three principal second-order NLO responses of few-layer PdX$_2$ materials,
namely, second-harmonic generation, linear electric-optical (LEO) effect and bulk photovoltaic effect.
The SHG, one of the best-known second-order NLO effects, converts the two photons of the same-frequency 
into a new photon with a doubled photon energy~\cite{Shen2003,Boyd2003}.
Noncentrosymmetric materials with a large SHG susceptibility $\chi^{(2)}(-2\omega; \omega,\omega)$ 
have many applications in nonlinear photonic devices such as frequency conversion. The
SHG is also a powerful probe of the symmetry of surfaces and interfaces~\cite{Shen2003}. 
Here we find that all PdX$_2$ structures possess large $\chi^{(2)}$ in the visible frequency spectrum 
and BL PdSe$_2$ has the largest $\chi^{(2)}$ value of 1.4$\times$10$^3$ pm/V at 1.9 eV.
The LEO effect refers to the linear refractive index change ($\Delta n$) 
with the applied electric field strength ($E$), $\Delta n = n^3rE/2$, where $n$ is
the refraction index and $r$ is the LEO coefficient \cite{Boyd2003}.
The LEO effect thus allows one to use an electrical signal to control the amplitude, 
phase or direction of a light beam in the NLO material,
and leads to a widely used means for high-speed optical modulation and sensing devices 
(see, e.g., Ref. \cite{Wu1996} and references therein).
We find that the LEO coefficients of BL PdS$_2$ and BL PdSe$_2$ are significant
and comparable to that of trilayer group 6B TMDC semiconductors~\cite{Wang2015b}.
The BPVE (also known as photogalvanic effect) refers to the generation 
of dc photocurrents in noncentrosymmetric materials~\cite{Sturman1992}. 
In a nonmagnetic semiconductor, there are two main contributions to the BPVE, 
namely, the circular injection current and linear shift current~\cite{Sturman1992,Sipe2000,Ahn2020,Gudelli2020}.
Materials having large BPVE are crucial for applications in photovoltaic solar cells and high sensitive photodetectors.
Here we predict that the BPVE in the considered few-layer PdX$_2$ structures is generally
strong, with a large shift current conductivity of up to 130 $\mu$A/V$^2$ 
and injection current susceptibility of up to 100$\times$10$^8$ A/V$^2$s in the visible frequency range.
These superior NLO responses of the BL and FL PdX$_2$ structures 
will make them valuable for technological applications in NLO and electro-optic 
devices such as light signal modulators, frequency converters, electro-optical switches, 
photovoltaics and photodetector applications. 

\section{Computational methods}
Bulk PdX$_2$ crystallize in a layered orthorhombic structure with space group $Pbca$ (see Fig. 1), as mentioned above.
The experimental lattice constants are $a$=5.460~\AA, $b$=5.541~\AA~and $c$=7.531~\AA~ for PdS$_2$\cite{Gronvold1957}
and $a$=5.7457~\AA, $b$=5.8679~\AA~and $c$=7.6976~\AA~ for PdSe$_2$~\cite{Soulard2004}.
The bulk unit cell contains two X-Pd-X monolayers stacked along the $c$-axis and each nearly squared 
inplane unit cell contains two chemical formulas (f.u.) (i.e., six atoms) (see Fig. 1).
In each X-Pd-X layer, interestingly, a Pd atom bonds with four chalcogen atoms, and
Pd and chalcogen atoms form a rare pentagonal structure [see Fig. 1(c)]~\cite{Oyedele2017,Liang2019}.
In the present calculations, a BL (FL) structure is constructed by cutting two (four) X-Pd-X layers out of the bulk crystal.  
The slab-superlattice approach is adopted with the separations of neighboring slabs being at least 15 \AA.
%The experimental bulk structural parameters are used. 
We notice that a number of the structural optimization 
calculations for bulk PdSe$_2$ using more than ten exchange-correlation functionals
have been carried out~\cite{Oyedele2017,Li2018,Kuklin2019,Sun2018}. 
The discrepancies between the experimental and theoretical lattice constants are large, 
varying from 2 \% all the way up to 20 \% depending strongly on the exchange-correlation
functional used~\cite{Oyedele2017,Sun2018}.
On the other hand, the experimental inplane lattice constants of BL PdSe$_2$ 
are only slightly larger than the corresponding lattice constants of bulk PdSe$_2$~\cite{Li2018}. 
Thus, we use the experimental bulk structural parameters in the present calculations.
%optimized inplane lattice constants of few-layer PdSe$_2$ were found to agree well with
%experimental bulk lattice constants.~\cite{Oyedele2017,Li2018,Sun2018,Kuklin2019} Therefore, 
We believe that using the experimental structural parameters of atomically thin PdX$_2$ would not significantly
change the calculated electronic and optical properties of BL and FL PdX$_2$ to be presented below. 
 
The electronic structure calculations are performed using the accurate projector augmented wave method~\cite{Kresse1999} 
as implemented in the Vienna ab-initio simulation package (VASP)~\cite{Kresse1996a,Kresse1996b}. 
For the exchange-correlation potential, we adopt the generalized gradient approximation (GGA) 
of Perdew-Burke-Ernzerhof parametrization~\cite{Perdew1996}. A large plane-wave energy cutoff 
of 400 eV is used throughout the calculations. The valence configuration of Pd atom is taken 
as 4d$^{9}$ 5s$^1$, S atom is 3s$^2$ 3p$^4$ and for Se atom it is 4s$^2$ 4p$^4$. A $k$-point 
mesh of 18 $\times$ 18 $\times$ 1 is used in the Brillouin zone (BZ) integrations 
for few-layer PdX$_2$ structures. All the calculations are performed within the 
scalar-relativistic projector augmented potentials, with the energy convergence 
up to 10$^{-6}$ eV between the successive iterations.

All the linear and NLO properties of the 2D PdX$_2$ structures are calculated from the self-consistent electronic band structures
within the linear response formalism with the independent-particle approximation (IPA). 
Specifically, we first calculate the imaginary part ($\varepsilon''(\omega)$) of the dielectric function due to 
direct interband transitions by using the Fermi golden rule \cite{Guo2004,Guo2005b},
\begin{equation}
\varepsilon''_{a}(\omega) = \frac{4\pi^2}{\Omega\omega^2}
\sum_{i\in\rm{VB},j\in\rm{CB}} \sum_{k} \omega_k|p_{ij}^{a}|^{2} \delta(\epsilon_{{\bf k}j}-\epsilon_{{\bf k}i}-\omega)
\end{equation}
where $\omega$ is the photon energy and $\Omega$ is the unit-cell volume. VB and CB denote 
the valence and conduction bands, respectively. 
The dipole transition matrix elements 
$p^a_{ij} = \braket {\bf{k}\it{j}|\hat{p}_a|\bf{k}\it{i}}$ are obtained from the self-consistent band
structures within the PAW formalism \cite{Adolph2001}. Here $\ket{\bf{k}\it{n}}$ 
is the $\it{n}$th Bloch state wave function with crystal momentum $\bf{k}$, 
and $a$ denotes the Cartesian component.
The obtained $\varepsilon''(\omega)$ is used to get the real part ($\varepsilon'(\omega)$) 
of the dielectric function by a Kramer-Kronig transformation \cite{Guo2004,Guo2005b},
\begin{equation}
\varepsilon'(\omega) = 1+\frac{2}{\pi} P \int _{0}^{\infty}\frac{\omega'(\varepsilon''(\omega')}{\omega^{'2}-\omega^{2}}d\omega', 
\end{equation}
where $P$ is the principal value of the integral.

%The BPVE refers to the generation of dc photocurrents, a second-order NLO response 
%of a noncentrosymmetric material 
%under the applied optical electromagnetic fields $E_{b}$ and $E_{c}$~\cite{VonBaltz1981,Sipe2000}.
For the BPVE, the dc photocurrent density along the $a$-axis in a noncentrosymmetric material
under the applied optical electric fields $E_{b}$ and $E_{c}$ may be written as ~\cite{Sipe2000,Nastos2010,Ahn2020} 
\begin{equation}
J_{a}(0)=\sum_{bc}\sigma_{abc}(0;\omega,-\omega)E_b(\omega)E_c(-\omega),
\end{equation}
where the photocurrent conductivity $\sigma_{abc}$ is a third-rank tensor~\cite{Sipe2000}.
For a nonmagnetic semiconductor, the dc photocurrent contains two main contributions, namely, 
the linear shift current~\cite{Sipe2000} and also the circular injection current~\cite{Nastos2010}.
That is, $\sigma_{abc} = \sigma_{abc}^{sh} + \sigma_{abc}^{inj}$.
%However, in a PT-symmetric magnetic materials, both the linear shift current and circular injection currents would be zero, instead they possess circular shift current and linear injection current \cite{Gudelli2020,Ahn2020}.
Within the length gauge formalism, the shift current conductivity $\sigma_{abc}^{sh}$ can be written in terms of
the interband position matrix element $r_{ij}^a$ and its momentum derivative $r_{ij;b}^a$ ~\cite{Sipe2000}.
%[see Eq. (57) in Ref. ~\cite{Sipe2000}]. 
By replacing $r_{ij}^a$ with $p_{ij}^a/i\epsilon_{ij}$ 
where $\epsilon_{ji}=(\epsilon _{\bf{k}j}-\epsilon _{\bf{k}i})$, one may obtain
\begin{eqnarray}
\fl\sigma_{abc}^{sh}(\omega)&=\frac{{\pi}}{\Omega}\sum_{\bf k}w_{\bf k}\sum_{i \in VB}\sum_{j\in CB}\frac{1}{\epsilon_{ij}^2} \sum_{l \ne i,j}Im\{\frac{p_{il}^a \langle p_{ji}^{b} p_{lj}^{c}\rangle}{2\epsilon_{il}} + \frac{p_{lj}^a \langle p_{ji}^{b} p_{il}^{c}\rangle}{2\epsilon_{jl}} \} \delta(\epsilon_{ji}-\omega),
\end{eqnarray}
where $\langle p^b_{jl} p^c_{li}\rangle = \frac{1}{2}(p^b_{jl} p^c_{li}+ p^b_{li} p^c_{jl})$. 

The injection current conductivity $\sigma_{abc}^{inj} = \tau \eta_{abc}$ where $\tau$ is the relaxation time of
photoexcited carriers and $\eta_{abc}$ is the injection current susceptibility, 
%~\cite{Nastos2010}.
%The $\eta_{abc}$ 
which can also be written in terms of $r_{ij}^a$ and $r_{ij;b}^a$ [see Eq. (31) in Ref. ~\cite{Nastos2010}]. 
Again, by substituting $r_{ij}^a$ with $p_{ij}^a/i\epsilon_{ij}$, one would get 
%where $\epsilon_{ji}=(\epsilon _{\bf{k}j}-\epsilon _{\bf{k}i})$, one would get
\begin{eqnarray}
\fl\eta_{abc}^{inj}(\omega)=\frac{{2\pi}}{\Omega}\sum_{\bf k}w_{\bf k}\sum_{i \in VB}\sum_{j\in CB}
\frac{ \Delta_{ji}^a Im[p_{ji}^{b} p_{ij}^{c}] }{\epsilon_{ij}^2 } \delta(\epsilon_{ji}-\omega).\nonumber \\
\end{eqnarray}

% and is given as follows,($J$)  the BPVE, which define as the difference of real-space positions 
%of Bloch electrons between valence band and conduction band, expressed in terms of current density ($J$) 
%as a response to the electromagnetic field ${\mathcal{E}}$\cite{VonBaltz1981,Sipe2000,Ibanez-Azpiroz2018},
%\begin{equation}
%J^a = 2\sigma^{abc}(0;\omega,-\omega)Re[\mathcal{E}_b(\omega) \mathcal{E}_c(\omega)]\\
%\end{equation}
%where $\sigma^{abc}$ is the third-rank tensor, known as shift current conductivity given as follows,
%\begin{eqnarray}
%\fl\sigma^{abc}&=\frac{-i\pi e^3}{4\hbar^2}\int[d{\bf k}]\sum_ {n,m}f_{nm}(r^b_{mn}r^{c;a}_{nm}+r^c_{mn}r^{b;a}_{nm})\nonumber\\
%&\times[\delta(\omega_{mn}-\omega)+\delta(\omega_{nm}-\omega)].
%\end{eqnarray}
%where $n$ and $m$ are bands index, ${\bf k}$ is the wave vector, $f_{nm}=f_n-f_m$is the differences between the occupation factors and $\hbar\omega_{nm}=E_m-E_n$ is the difference between band energies and the integral is over the first BZ with [d${\bf k}$] = d$^d$/(2$\pi$)$^d$ in dimension $d$. The interband dipole matrix represented by $r^a_{nm}=(1-\delta_{nm})A^a_{nm}$, where $A^a_{nm} = i\braket{u_m|\partial_a u_n}$ is the Berry connection matrix.

The imaginary part of the SHG susceptibility 
%due to direct interband transitions 
is given by \cite{Guo2004,Guo2005a}
\begin{eqnarray}
\fl\chi''^{(2)}_{(abc)}(-2\omega,\omega,\omega) &= \chi''^{(2)}_{(abc,\rm{VE})}(-2\omega,\omega,\omega) 
+\chi''^{(2)}_{(abc,\rm{VH})}(-2\omega,\omega,\omega),
\end{eqnarray}
where the contribution due to the so-called virtual-electron (VE) process is \cite{Guo2004,Guo2005a}
\begin{eqnarray}
\fl\chi''^{(2)}_{(abc,\rm{VE})}&=-\frac{\pi}{2\Omega} \sum _{i\in \rm{VB}}\sum _{j,l\in \rm{CB}}\sum _{\bf{k}} \omega _{\bf{k}}\Bigg\{\frac{\rm{Im}[p^a_{jl}\langle p^b_{li} p^c_{ij}\rangle]}{\epsilon^3_{li}(\epsilon _{li}+\epsilon _{ji})} 
\times \delta(\epsilon _{li}-\omega)-\frac{\rm{Im}[p^a_{ij}\langle p^b_{jl} p^c_{li}\rangle]}{\epsilon^3_{li}(2\epsilon _{li}-\epsilon _{ji})}\delta(\epsilon _{li}-\omega) \nonumber\\
&+\frac{\rm{16Im}[p^a_{ij}\langle p^b_{jl} p^c_{li}\rangle]}{\epsilon^3_{ji}(2\epsilon^3_{li}-\epsilon^3_{ji})}\delta(\epsilon _{ji}-2\omega)\Bigg\}
\end{eqnarray}
and that due to the virtual-hole (VH) process is \cite{Guo2004,Guo2005a}
\begin{eqnarray}
\fl\chi''^{(2)}_{(abc,\rm{VH})}&=\frac{\pi}{2\Omega}\sum _{i,l\in\rm{VB}}\sum _{j\in\rm{CB}}\sum _{\bf{k}} \omega _{\bf{k}}\Bigg\{\frac{\rm{Im}[p^a_{li}\langle p^b_{ij} p^c_{jl}\rangle]}{\epsilon^3_{jl}(\epsilon _{jl}+\epsilon _{ji})}  
\times \delta(\epsilon _{jl}-\omega)-\frac{\rm{Im}[p^a_{ij}\langle p^b_{jl} p^c_{li}\rangle]}{\epsilon^3_{jl}(2\epsilon _{jl}-\epsilon _{ji})}\delta(\epsilon _{jl}-\omega) \nonumber\\
&+\frac{\rm{16Im}[p^a_{ij}\langle p^b_{jl} p^c_{li}\rangle]}{\epsilon^3_{ji}(2\epsilon^3_{jl}-\epsilon^3_{ji})}\delta(\epsilon _{ji}-2\omega)\Bigg\}.
\end{eqnarray}
%Here $\epsilon_{ji}=(\epsilon _{\bf{k}j}-\epsilon _{\bf{k}i})$ and 
%Here $\langle p^b_{jl} p^c_{li}\rangle = \frac{1}{2}(p^b_{jl} p^c_{li}+ p^b_{li} p^c_{jl})$. 
The real part of the SHG susceptibility is then obtained from $\chi'^{(2)}_{abc}$ by the Kramer-Kronig transformation \cite{Guo2004,Guo2005a}:
\begin{equation}
\chi'^{(2)}_{abc}(-2\omega,\omega,\omega) = \frac{2}{\pi} P \int _{0}^{\infty}\frac{\omega'(\chi''^{(2)}(2\omega',\omega',\omega')}{\omega^{'2}-\omega^{2}}d\omega'.
\end{equation}

We also calculate the low-frequency LEO coefficients of the considered materials 
using the obtained static dielectric constants and SHG susceptibility. 
%The LEO effect is considered as a special effect of the second-order NLO properties. 
%The LEO coefficients of few-layer PdX$_2$ structures 
The LEO coefficients in the zero frequency limit are given by
%in terms of the second-order optical susceptibility of $\chi''^{(2)}_{abc}(-\omega,\omega,0)$ as
\begin{equation}
r_{abc}(0) = -\frac{2}{\varepsilon _a(0)\varepsilon _b(0)} \lim_{\omega\to 0} \chi''^{(2)}_{abc}(-2\omega,\omega,\omega).
\end{equation}

To ensure the accuracy of the calculated optical properties we use denser $k$-point meshes 
of 110$\times$110$\times$1 for BL structures and 60$\times$60$\times$1 for FL structures. 
The $\delta$ function in Eqs. 4, 5, 7 and 8 is approximated 
by a Gaussian function with $\Gamma = 0.1$ eV. Furthermore, to ensure that $\epsilon'(\omega)$ 
and $\chi'^{(2)}$ calculated via Kramer-Kronig transformation Eqs. 2 and 9 are reliable, 
at least 150 and 300 energy bands are included in the present optical calculations for BL and FL structures, 
respectively. The unit-cell volume $\Omega$ in Eqs. 4, 5, 7 and 8 is not well-defined for 
low-dimensional systems. Therefore, similar to the previous 
calculations \cite{Guo2004,Guo2005a,Guo2005b,Wang2015b}, we use the effective unit-cell volume
(i.e., $\Omega_{eff}=a\times b\times nd$, where $n$ is the layer-number and $d$ is the effective layer thickness)  
of the 2D material rather than the volume of the supercell which is arbitrary.
Since each unit cell of bulk PdX$_2$ contains two layers, the effective layer thickness $d$ 
is 3.766 \AA$ $ for few-layer PdS$_2$ and 3.849 \AA$ $ for few-layer PdSe$_2$. 

Equations (1), (4), (5), (7) and (8) indicate that correct band gaps 
would be important for obtaining accurate optical properties. 
However, in general the GGA functional is known to underestimate the band gaps because 
some many-body effects especially the quasiparticle self-energy corrections are neglected.
Therefore, we perform the band-structure calculations using the hybrid Heyd-Scuseria-Ernzerhof (HSE) functional, 
which is known to produce much improved band gaps for semiconductors\cite{Heyd2003,Heyd2006}. 
We then use the HSE band gaps and calculate all the optical properties with the well-known scissors-correction 
(SC) \cite{Levine1991}. In the SC calculations, the conduction bands are uniformly up-shifted 
so that the band gap would match the HSE gap together with the renormalized transition 
matrix elements \cite{Levine1991}. All the optical properties presented in this paper
are obtained with this SC scheme.

\begin{figure}[htb]
\begin{center}
\includegraphics[width=13cm]{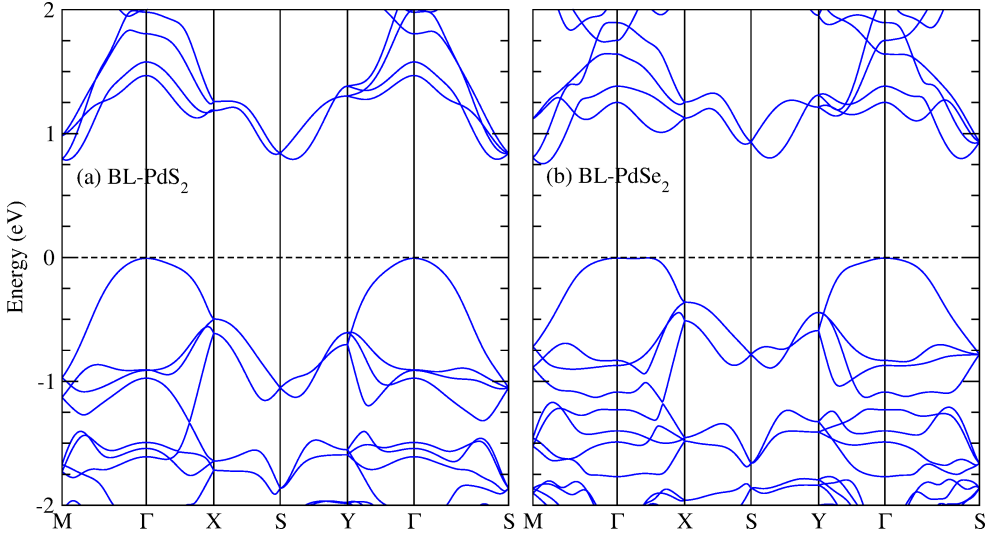}
\caption{Band structures of (a) BL PdS$_2$ and (b) BL PdSe$_2$. The horizontal dashed 
line denotes the top of valance bands.}
\end{center}
\label{band}
\end{figure}

\section{Results and discussion}
\subsection{Electronic structure}
The calculated GGA band structures of BL PdX$_2$ and FL PdX$_2$ are presented, respectively,
in Fig. 2 and Fig. S1 of the supplementary information (SI). 
%Since the band-structure profiles of 
%FL PdX$_2$ structures are very similar to that of the BL structures we presented them in Fig. S1 
%of the supplementary information (SI). 
All the four structures are an indirect band gap semiconductor.
% with indirect band gaps. 
The valence band maximum (VBM) is located at the $\Gamma$ point and 
the conduction band minimum (CBM) is located on the M-$\Gamma$ line in the 2D Brillouin zone
[see Fig. 1(d)] for both BL 
and FL PdS$_2$ structures [Fig. 2(a) and Fig. S1(a)]. 
In the PdSe$_2$ structures, the VBM is located in the $\Gamma$-X direction and the CBM is located 
on the M-$\Gamma$ line [Fig. 2(b) and Fig. S1(b)]. 
%making these structures as an indirect band gap semiconductors, in agreement with 
The present band structure of BL PdSe$_2$ agrees rather well with that reported 
in Refs. \cite{Oyedele2017,Kuklin2019,Li2018}. 
%The BL and FL PdSe$_2$ as well as ML and bulk materials of PdSe$_2$ are all indirect 
%band gap \cite{Sun2015,Oyedele2017,Kuklin2019}, in contrast to Mo-based TMDCs where 
%indirect-direct band gap transition was observed when moving from bulk to ML.
The calculated band gaps of all the structures are listed in Table 1,
together with the available experimental band gaps for BL and FL PdSe$_2$ ~\cite{Li2018,Oyedele2017}.
Table 1 shows that the band gap decreases significantly 
as the S atoms are replaced by the Se atoms in PdX$_2$ 
and also as we move from the BL to FL PdX$_2$ structure, 
indicating the tunability of the band gap by chalcogen substitution and also by layer-number variation. 
%represents layer dependent band gaps in PdX$_2$ structures. 

\begin{figure}[htb]
\begin{center}
\includegraphics[width=13cm]{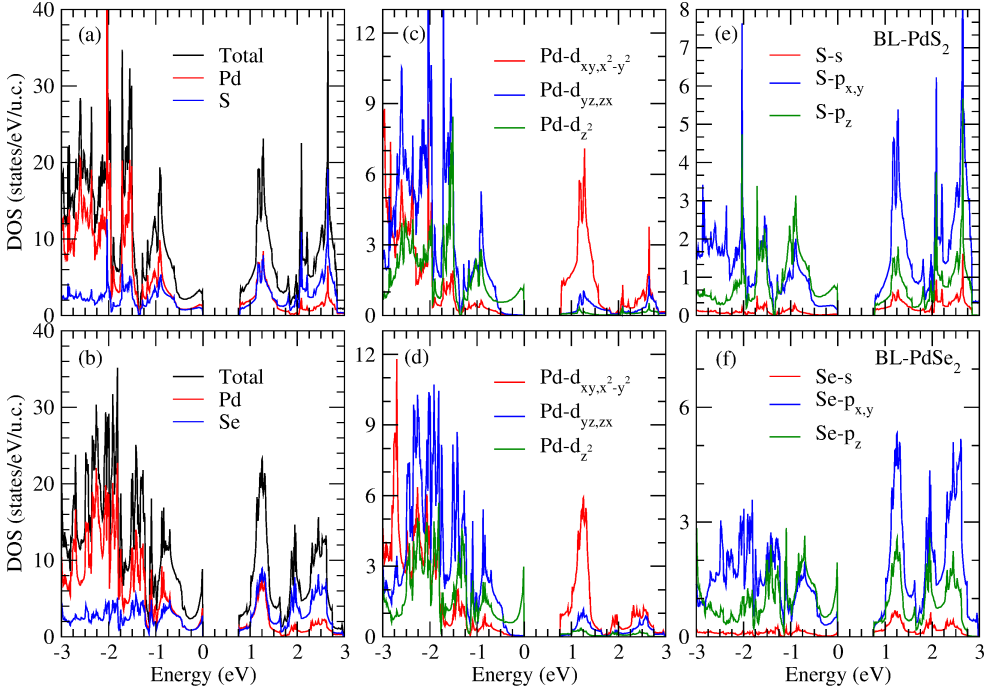}
\end{center}
\caption{Total- and orbital-projected densities of states (DOS) of BL PdS$_2$ (upper panels)
and BL PdSe$_2$ (lower panels).}
\label{dos}
\end{figure}

%We have calculated the total- and orbital-projected density of states (DOS) for BL and FL PdX$_2$ structures.
The calculated total- and orbital-projected density of states (DOS) for BL PdX$_2$ and FL PdX$_2$ 
are presented in Fig.~\ref{dos} and in Fig. S3 of the SI, respectively.
%The total and orbital projected DOS spectra obtained for PdS$_2$ and PdSe$_2$ FL structures
%are presented in Fig. S3 of SI, are rather similar to those of PdS$_2$ and PdSe$_2$ BL structures (Fig. 3).
%However, the states in the FLs are much higher compared to the BL structures. 
Figure~\ref{dos} shows that the upper valence band edge and lower conduction band edge
are contributed almost equally by the Pd $d$ orbitals and chalcogen (X) $p$ orbitals [see Figs.~\ref{dos}(a) and \ref{dos}(b)]. 
This indicates a strong covalent bonding in the PdX$_2$ structures (see also ~\cite{Guo1986}), being in rather
strong contrast to group 6B TMDCs (e.g., MoS$_2$) which may be called charge-transfer semiconductors~\cite{Wang2015b}.
This is due to the nearly filled Pd $d$ states in the PdX$_2$ structures
while in MoS$_2$ the Mo $d$ states are less than half-filled~\cite{Wang2015b}.
%that the presence of hybridization between the Pd $d$ states and chalcogen $p$ states results in
%a strong covalent bonding in both BL and FL PdX$_2$ structures, which is similar to that of the other TMDCs.
%However, in the present case, the hybridization among the Pd and chalcogen atoms is very strong compared
%to the other TMDCs (e.g., MoS$_2$), because of the nearly equal contribution of DOS from both
%the atoms at both the band edges (see Fig. \ref{dos}(a), (b)). Moreover, the nearly filled $d$ orbital
%of the Pd atom enhances the hybridization strength compared to partially filled Mo $d$ orbital of
%MoS$_2$, this leads to a strong interlayer coupling in PdX$_2$ \cite{Oyedele2017}. 
Furthermore, orbital-projected DOS spectra show that the contribution at the upper valence band edge 
comes predominantly from Pd $d_{z^2}$ and chalcogen $p_z$ with minor contribution from $p_{x,y}$ states 
while the contribution at the lower conduction band edge
comes from Pd $d_{xy,x^2-y^2}$ and chalcogen $p_{x,y}$ and $p_z$ states. Thus, the optical transitions
in PdX$_2$ take place from the valence states of the hybridized Pd $d$ and chalcogen $p$ states
to the conduction band states of Pd $d$ orbitals. 
%In addition, the sharp increase in the DOS
%at the valence band edge, due to the presence of a quasi-flat valence band along the $\Gamma$-X direction
%of the BZ (see Fig. \ref{band}(b)), suggests that PdSe$_2$ may produce a higher value
%of the Seebeck coefficient similar to the ML PdSe$_2$ \cite{Sun2015,Lan2019}. 
%The higher DOS
%at the valence band edge shows a strong SHG susceptibility, which will be explained in the
%following section \cite{Stoumpos2015}.

Table 1 shows that the GGA band gaps for BL and FL PdSe$_2$ 
are significantly smaller than the corresponding experimental values \cite{Li2018,Oyedele2017}, 
indicating that the GGA functional generally underestimates 
the band gaps of semiconductors, as mentioned in the preceding section. 
%underestimated for all the structures. 
%The obtained GGA functional band gap of the BL (FL) PdSe$_2$ is 0.74 (0.27) eV is underestimated 
%in comparison with the measured experimental band gap of 1.15 (1.06) eV \cite{Li2018,Oyedele2017}. 
%There are no experimental reports are available to compare the band gaps of both BL and FL PdS$_2$ structures.
Therefore, we also perform the band-structure calculations using the HSE functional~\cite{Heyd2003,Heyd2006}
and the HSE band structures are presented in Fig S2 in the SI.
%Clearly, the GGA functional is known for underestimating the band gaps.
%The calculated band-structure using the HSE functional are presented in Fig S2. 
Although the dispersions of the HSE band structures are similar to that of the corresponding GGA band structures,
% profiles and the band gap nature of the HSE are quite similar to 
%that of the GGA band-structures, except 
the band gaps from the HSE functional are significantly larger than that of the GGA ones
(Table 1). Furthermore, the HSE band gaps of BL and FL PdSe$_2$ are in better 
agreement with the experimental values than the GGA band gaps (Table 1).
Therefore, all the optical properties presented in the following sections 
are calculated within the scissors-correction scheme~\cite{Levine1991} by using the HSE band gaps (Table 1).
%magnitude of the band gap is improved in the HSE. 
%With HSE functionals, as mentioned the band gap values of all the structures are enhanced and 
%are presented in Table 1. However the band gaps using the HSE functional are found to be 
%in good agreement with the experimental reports of BL and FL PdSe$_2$ structures, 
%albeit a little higher is found for BL PdSe$_2$. 

\begin{table*}[htbp]
\begin{center}
\caption{Band gaps (in eV) of BL and FL PdX$_2$ calculated  with GGA and HSE functional. 
The available experimental band gap values are in parentheses for comparison.}
\begin{tabular}{cccccc}
\hline \hline
&					PdS$_2$		&&	&PdSe$_2$		\\
\hline	
					&GGA	&HSE		&	&GGA	&HSE		\\
\hline	
BL					&0.78	&1.90 		&	&0.74	&1.62 (1.15\cite{Li2018})	\\
$\triangle E=E_{HSE}$-$E_{GGA}$	&	&1.12		&	&	&0.88			\\
FL					&0.40	&1.47		&	&0.27	&1.11 (1.06\cite{Oyedele2017})	\\
$\triangle E=E_{HSE}$-$E_{GGA}$	&	&1.07		&	&	&0.84				\\
\hline \hline
\end{tabular} \\
\end{center}
\label{table1}
\end{table*}

\subsection{Linear optical properties}
\begin{figure}[htb]
\begin{center}
\includegraphics[width=13cm]{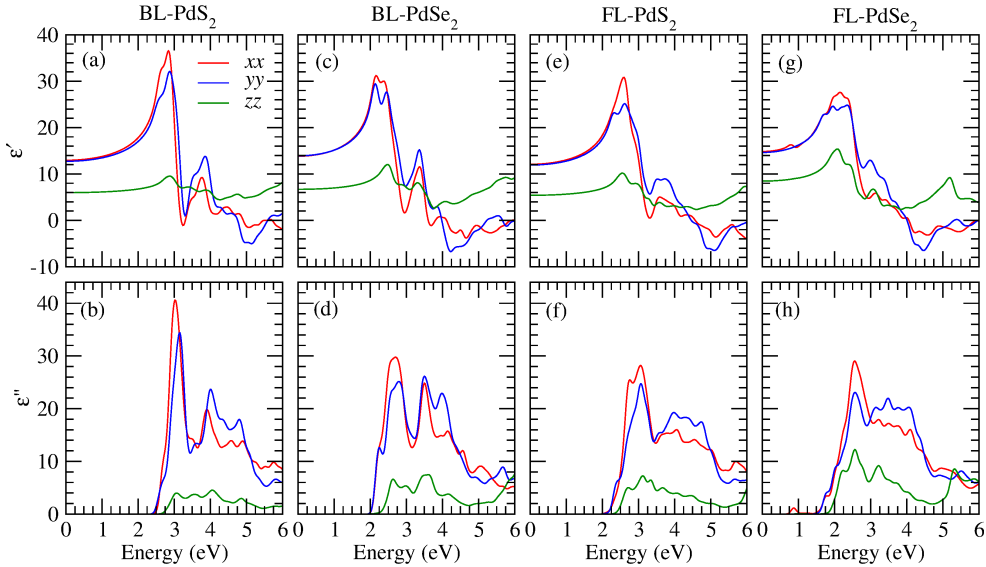}
\end{center}
\caption{Real (upper panels) and imaginary (lower panels) parts of the dielectric functions of BL-PdS$_2$, 
BL-PdSe$_2$, FL-PdS$_2$, and FL-PdSe$_2$.}
\label{linear}
\end{figure}

%In this section we present the linear optical properties of the BL and FL PdX$_2$ structures.
%real ($\varepsilon'(\omega)$) and imaginary ($\varepsilon''(\omega)$) 
%part of the optical dielectric function of PdX$_2$ BL and FL structures. 
The calculated dielectric functions of all the four PdX$_2$ structures are plotted in Fig.~\ref{linear}.
%linear optical properties 
%of these structures are presented in Fig.~\ref{linear}. To address the anisotropy we have studied 
%the optical dielectric function in different directions, that is, $\varepsilon_{xx}$, $\varepsilon_{yy}$ 
%and $\varepsilon_{zz}$. 
As can be expected of a 2D material, there are huge differences between 
the out-of-plane and in-plane components of the dielectric functions (see Fig.~\ref{linear}). 
For example, the real part of the $z$-polarized dielectric function $\varepsilon'_{zz}$ is only
about half of that for the in-plane polarized dielectric function $\varepsilon'_{xx}$ and $\varepsilon'_{yy}$ below 3 eV.
%dielectric tensors of $\varepsilon'_{xx}$ and $\varepsilon'_{yy}$ below 3 eV and above this energy
%the $\varepsilon'_{zz}$ if found to be little higher in value compared to the other two components.
The imaginary part of the $z$-polarized dielectric function $\varepsilon''_{zz}$ is about four times smaller
than that of the in-plane dielectric functions $\varepsilon''_{xx}$ and $\varepsilon''_{yy}$.
%In the case of imaginary dielectric tensor, the $\varepsilon''_{zz}$ is much smaller in value compared
%to the other two dielectric tensors of $\varepsilon''_{xx}$ and $\varepsilon''_{yy}$.
%Although much smaller, 
There are also discernable differences between the two in-plane components
of the dielectric functions, namely, $\varepsilon_{xx}$ and $\varepsilon_{yy}$ (see Fig.~\ref{linear}).
%In Fig.~\ref{linear}, large anisotropy is evident in both optical dielectric 
%tensors for all the structures. 
%For all the structures the real optical dielectric functions of $\varepsilon'_{xx}$ 
%and $\varepsilon'_{yy}$ are found to be similar except at few energy ranges. 
In particular, the real dielectric constant of $\varepsilon'_{xx}$ is slightly higher than $\varepsilon'_{yy}$ 
in the energy range of 2.5-3.0 (2.0-2.5) eV and 4.3-5.6 (4.0-5.0) eV, whereas $\varepsilon'_{yy}$ 
in the energy range of 3.0-4.3 (2.5-4.0) eV and also in 5.6-6.0 (5.0-5.7) eV is higher 
than $\varepsilon'_{xx}$ for BL PdS$_2$ (PdSe$_2$) (see Fig.~\ref{linear}(a) and (c)). 
In FL PdS$_2$ (PdSe$_2$) structure, the real dielectric constant $\varepsilon'_{xx}$ is 
higher than $\varepsilon'_{yy}$ in the energy range of 2.2-2.7 (1.9-2.5) eV and 4.3-5.6 (4.0-5.6) eV, 
whereas $\varepsilon'_{yy}$ in the energy range of 2.7-4.3 (2.5-4.0) eV and also 
in 5.6-6.00 (5.0-6.0) eV is slightly higher than $\varepsilon'_{xx}$ (see Fig.~\ref{linear}(e) and (g)). 
Similar profile is found in the imaginary part of the dielectric constant, where $\varepsilon''_{xx}$ 
is higher than $\varepsilon''_{yy}$ in the energy range of 2.6-3.1 eV (2.5-3.4 eV) 
and 3.9-5.0 eV (5.0-6.0 eV) for BL (FL) PdS$_2$ and 2.2-2.9 eV (2.1-3.0 eV) and 4.5-5.4 eV (4.5-5.3 eV) 
for BL (FL) PdSe$_2$, whereas in the other energy windows $\varepsilon''_{yy}$ is higher 
than $\varepsilon''_{xx}$ for both BL (FL) PdS$_2$ and PdSe$_2$ structures, see Fig.~\ref{linear}(b) 
and (d) (Fig.~\ref{linear}(f) and (h)). 
%Further, the real dielectric tensor of of $\varepsilon'_{zz}$ is almost half of the other 
%dielectric tensors of $\varepsilon'_{xx}$ and $\varepsilon'_{yy}$ below 3 eV and above this energy 
%the $\varepsilon'_{zz}$ if found to be little higher in value compared to the other two components. 
%In the case of imaginary dielectric tensor, the $\varepsilon''_{zz}$ is much smaller in value compared 
%to the other two dielectric tensors of $\varepsilon''_{xx}$ and $\varepsilon''_{yy}$. 
%The above analysis clearly shows that few-layers PdX$_2$ structures are highly anisotropic in nature. 
%Further, the imaginary part of the dielectric function allows us to the information regarding single 
%and double photons contribution to the SHG susceptibility, which will be discussed in details 
%in the next section.
It is worth mentioning that the large imaginary part of the dielectric function of all the PdX$_2$ structures
spans over a wide range of the visible frequency range, which is similar to that of silicon~\cite{Albrecht1998}.
This suggests that as for silicon, these 2D PdX$_2$ structures will be useful for opto-electronic applications
such as high solar-absorption efficiency solar cells~\cite{Cai2017}.

%More advanced {\it ab initio} many-body theoretical calculation (namely, so-called GW$_0$-BSE calculation)
%of the absorptive part of the dielectric function of BL PdSe$_2$ has been reported~\cite{Kuklin2019}. 
%%We compare linear optical properties of the imaginary part of the dielectric constant 
%%of BL PdSe$_2$ structure with the most accurate GW$_0$ derived quasi-particle band gap dielectric 
%%constant approximated by Bethe-Salpeter equation (GW$_0$-BSE) \cite{Kuklin2019}.
%It is rather encouraging to note that overall, the imaginary part of the present
%dielectric function of BL PdSe$_2$ agrees quite well with that obtained 
%from the GW$_0$-BSE calculation~\cite{Kuklin2019}.
%constant from the experimental
%band gap and GW$_0$-BSE reported results are in good agreement with each other.
%In particular, the first optical transition occurs at photon energy
%of 2.1 eV, being in good agreement with that from the GW$_0$-BSE calculation at 1.89 eV.

\subsection{Shift and injection currents}
%In this section, we present the calculated shift and injection currents due to the BPVE in BL 
%and FL PdX$_2$. 
As mentioned before, the point symmetry group of BL and FL PdX$_2$ is
$C_{2v}$ with the $C_2$ rotation axis along the $y$-axis. Therefore,
there are only five inequivalent nonzero shift current tensor elements~\cite{Gallego2019}, 
namely, $\sigma^{sh}_{xxy}=\sigma^{sh}_{xyx}$, $\sigma^{sh}_{yxx}$, 
$\sigma^{sh}_{yyy}$, $\sigma^{sh}_{yzz}$ and $\sigma^{sh}_{zyz}=\sigma^{sh}_{zzy}$.
Similarly, there are only two inequivalent nonzero injection current susceptibility elements,
namely, $\eta_{xxy} = -\eta_{xyx}$ and $\eta_{zzy} = -\eta_{zyz}$~\cite{Gallego2019}.
Since the photocurrent cannot flow along the out-of-plane direction (the $z$-axis),
we will not consider the nonzero elements of $\sigma^{sh}_{zyz}=\sigma^{sh}_{zzy}$ and $\eta_{zzy} = -\eta_{zyz}$ 
in the rest of this paper. 
  
\begin{figure}[htb]
\begin{center}
\includegraphics[width=13cm]{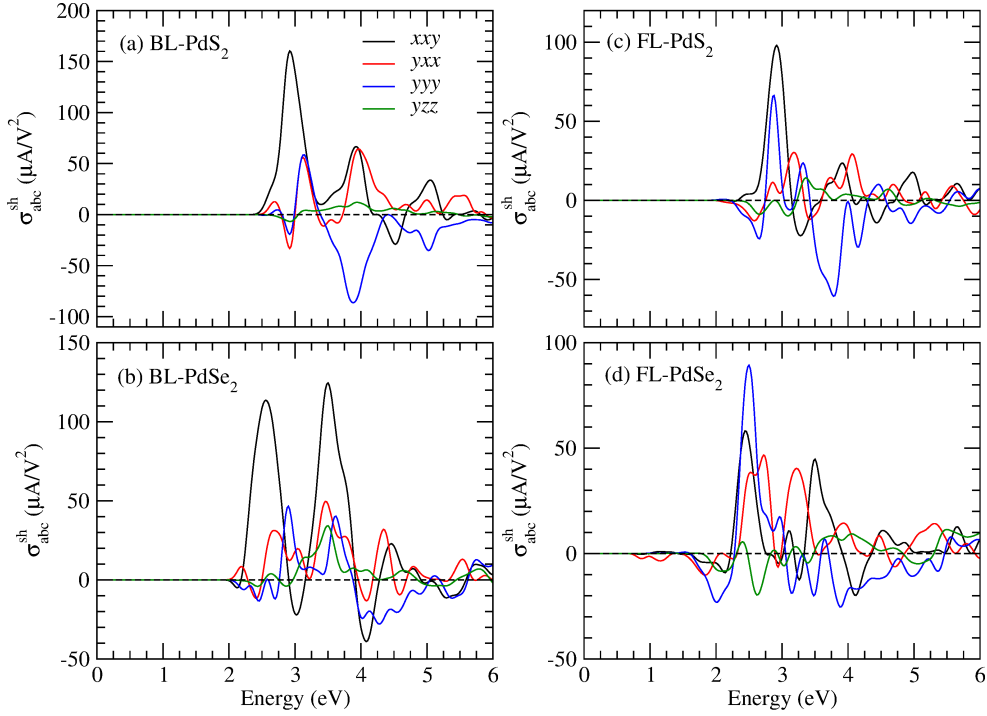}
\end{center}
\caption{The shift current conductivity of (a) BL PdS$_2$, (b) BL PdSe$_2$, (c) FL PdS$_2$ 
and (d) FL PdSe$_2$.}
\label{shiftc}
\end{figure}

%i.e., $\sigma^{sh}_{xxy}$, $\sigma^{sh}_{xyx}$, $\sigma^{sh}_{yxx}$, $\sigma^{sh}_{yyy}$, 
%$\sigma^{sh}_{yzz}$, $\sigma^{sh}_{zyz}$ and $\sigma^{sh}_{zzy}$, where the subscripts $x$, $y$ 
%and $z$ represent the three Cartesian coordinates. Taking symmetry into  account only five 
%independent elements remains, namely $\sigma^{sh}_{xxy}=\sigma^{sh}_{xyx}$, $\sigma^{sh}_{yxx}$, 
%$\sigma^{sh}_{yyy}$, $\sigma^{sh}_{yzz}$ and $\sigma^{sh}_{zyz}=\sigma^{sh}_{zzy}$. 
%The considered symmetry is consistent with our theoretical results therefore showing that 
%our numerical calculations are qualitatively correct. 
%Among the above five independent elements we focus only on four elements by discarding $\sigma^{sh}_{zyz}$, 
%since $z$ is the out-of-plane direction.
The calculated four inequivalent nonzero shift current conductivity elements 
for BL and FL PdX$_2$ are plotted in Fig. \ref{shiftc}. 
Figure \ref{shiftc} shows that in all the PdX$_2$ structures, the four shift current conductivity elements
are zero below the band gap but increase rapidly above the band gap.
Notably, the $\sigma^{sh}_{xxy}$ element in BL PdS$_2$ 
dominates the low photon energy range of 2.4$\sim$3.2 eV with a prominent peak
of height of 160 $\mu$A/V$^2$ at 2.9 eV. The $\sigma^{sh}_{xxy}$ of BL PdS$_2$ has
a second peak with the reduced maximum of 65 $\mu$A/V$^2$ at $3.9$ eV [see Fig. \ref{shiftc}(a)].
For BL PdSe$_2$, these two prominent peaks in the $\sigma^{sh}_{xxy}$ spectrum
become comparable with the maximum values of $\sim$115 $\mu$A/V$^2$ and $\sim$126 $\mu$A/V$^2$
at 2.6 eV and 3.6 eV, respectively [see Fig. \ref{shiftc}(b)]. 
In both BL structures, the magnitudes of the $\sigma^{sh}_{yxx}$ and $\sigma^{sh}_{yyy}$ 
spectra are also pronounced in the photon energy range from the absorption edge to $\sim$5.0 eV. 
For example, the $\sigma^{sh}_{yyy}$ of BL PdS$_2$
has a negative peak at $\sim$3.9 eV with the maximum value of -90 $\mu$A/V$^2$.
% and of from $\sigma^{sh}_{xxy}$ at a photon energy of 2.9 eV. 
%The second maxima value of $\sim$-90 $\mu$A/V$^2$ is found for PdS$_2$ BL from $\sigma^{sh}_{yyy}$ 
%at a photon energy of $\sim$3.9 eV (see Fig. \ref{shiftc}(a)). In the case of PdSe$_2$ 
%BL structure a maximum value of the order of $\sim$126 $\mu$A/V$^2$ from $\sigma^{sh}_{xxy}$ 
%at a photon energy of 3.5 eV and a second maximum from the same element of $\sim$115 $\mu$A/V$^2$ 
%at $\sim$2.6 eV (see Fig. \ref{shiftc}(b)). Other shift current conductivity elements, 
%except $\sigma^{sh}_{yzz}$, also have a maximum value of about $\sim$50 $\mu$A/V$^2$ 
%in the energy range of 3-4 eV for both PdS$_2$ and PdSe$_2$ BL structures (see Fig. \ref{shiftc}(a) and (b)). 
Fig. \ref{shiftc} also indicates that the magnitude of the $\sigma^{sh}_{yzz}$ spectra from all the four structures
is much smaller than all the other shift conductivity elements. This may be attributed to
the fact that the absorptive part of the out-of-plane polarized dielectric function element $\varepsilon_{zz}$ 
is much smaller than that of the in-plane polarized dielectric function elements $\varepsilon_{xx}$ and $\varepsilon_{yy}$
(see Fig. \ref{linear}). 
In the FL structure, the maximum shift current conductivity is 
from $\sigma^{sh}_{xxy} $($\sigma^{sh}_{yyy}$) of the order of $\sim$97 ($\sim$90) $\mu$A/V$^2$ 
at a photon energy of 2.9 (2.5) eV for PdS$_2$ (PdSe$_2$) [see Fig. \ref{shiftc}(c) and (d), 
respectively]. Similar to the BL structures, FL PdX$_2$ also have contributions of $\sim$50 $\mu$A/V$^2$ 
from other elements such as $\sigma^{sh}_{yyy}$ for FL PdS$_2$ and $\sigma^{sh}_{xxy}$, 
$\sigma^{sh}_{yxx}$ for FL PdSe$_2$ in the energy range of 3-4 eV [see Fig. \ref{shiftc}(c) and (d)]. 
%The observed shift current response of the BPVE from both BL and FL PdX$_2$ structures 
%is found to be very high compared to the known BPVE materials \cite{Young2012,Rangel2017,Cook2017}.

\begin{figure}[htb]
\begin{center}
\includegraphics[width=13cm]{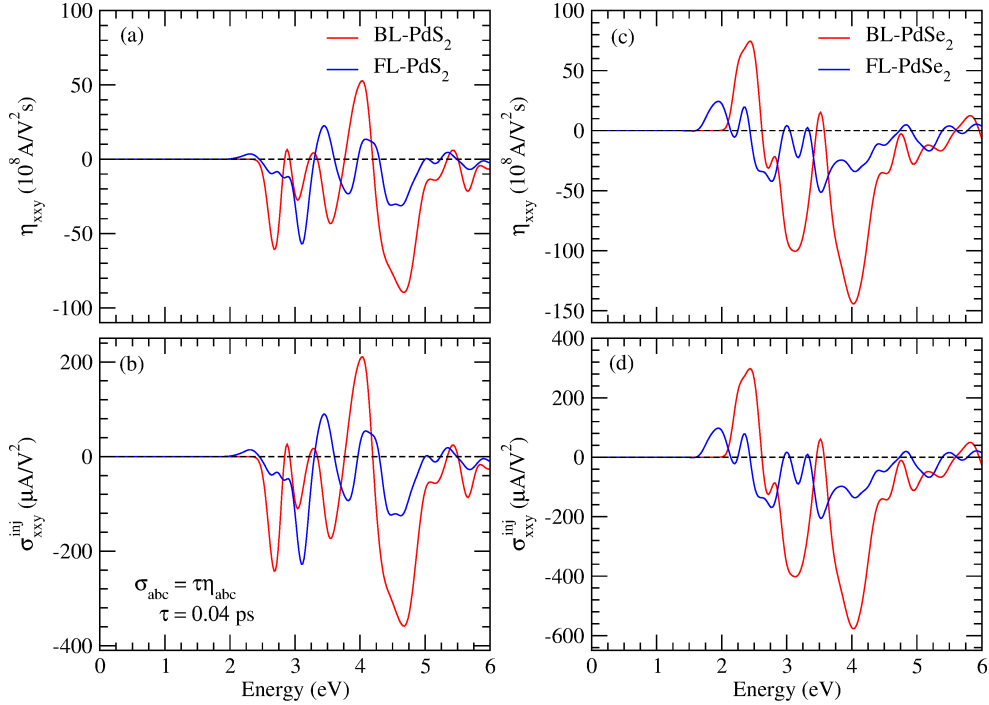}
\end{center}
\caption{(a) [(c)] Injection current susceptibility ($\eta_{xxy}$) tensor and (b) [(d)] injection conductivity 
($\sigma_{xxy}$) of BL and FL PdS$_2$ [PdSe$_2$].}	
\label{injection}
\end{figure}

The calculated nonzero injection current susceptibility element $\eta_{xxy}$ is
displayed as a function of photon energy for both BL and FL PdX$_2$ in Fig. \ref{injection}.
Using the typical relaxation time $\tau$ = 0.04 ps for 2D PdX$_2$ materials\cite{Lan2019}, 
we obtain the injection photocurrent conductivity 
$\sigma^{inj}_{xxy} = \tau \eta_{xxy}$, as shown in Fig. \ref{injection}. 
For all the PdX$_2$ structures, both $\eta_{xxy}$ and $\sigma^{inj}_{xxy}$ 
are zero below the absorption edge but they increase rapidly above the absorption edge. 
In BL PdS$_2$, the calculated injection conductivity $\sigma^{inj}_{xxy}$ 
has two negative prominent peaks with the maximum values of -240 
and -360 $\mu$A/V$^2$ at 2.7 and 4.7 eV, respectively [see Fig. \ref{injection}(b)]. 
Similarly, in BL PdSe$_2$, we also find two pronounced peaks with larger maximum values 
of -400 and -580 $\mu$A/V$^2$ at 3.1 and 4.0 eV, respectively [see Fig. \ref{injection}(d)]. 
In the FL structures, the magnitudes of both $\eta_{xxy}$ and $\sigma^{inj}_{xxy}$
spectra are smaller compared with that from the BL structures
(Fig. \ref{injection}). In particular, the magnitudes of the $\eta_{xxy}$ and $\sigma^{inj}_{xxy}$
of FL PdSe$_2$ are generally less than half of that from BL PdSe$_2$ 
[see Figs. \ref{injection}(c) and \ref{injection}(d)]. 
Nonetheless, FL PdS$_2$ and FL PdSe$_2$ do have a rather pronounced peak
in the $\sigma^{inj}_{xxy}$ spectrum with the peak value being  
about -200 $\mu$A/V$^2$ at 3.1 and 2.8 eV, respectively.
%The injection currents of all the four structures appear to be nearly four-times higher 
%in magnitude compared to the shift currents, this shows that the injection current is dominated in 
%the PdX$_2$ materials. Nonetheless, the size of the shift currents as well as injection currents 
%are comparable and even high compared to the other known BPVE materials. 

\begin{figure}[htb]
\begin{center}
\includegraphics[width=13cm]{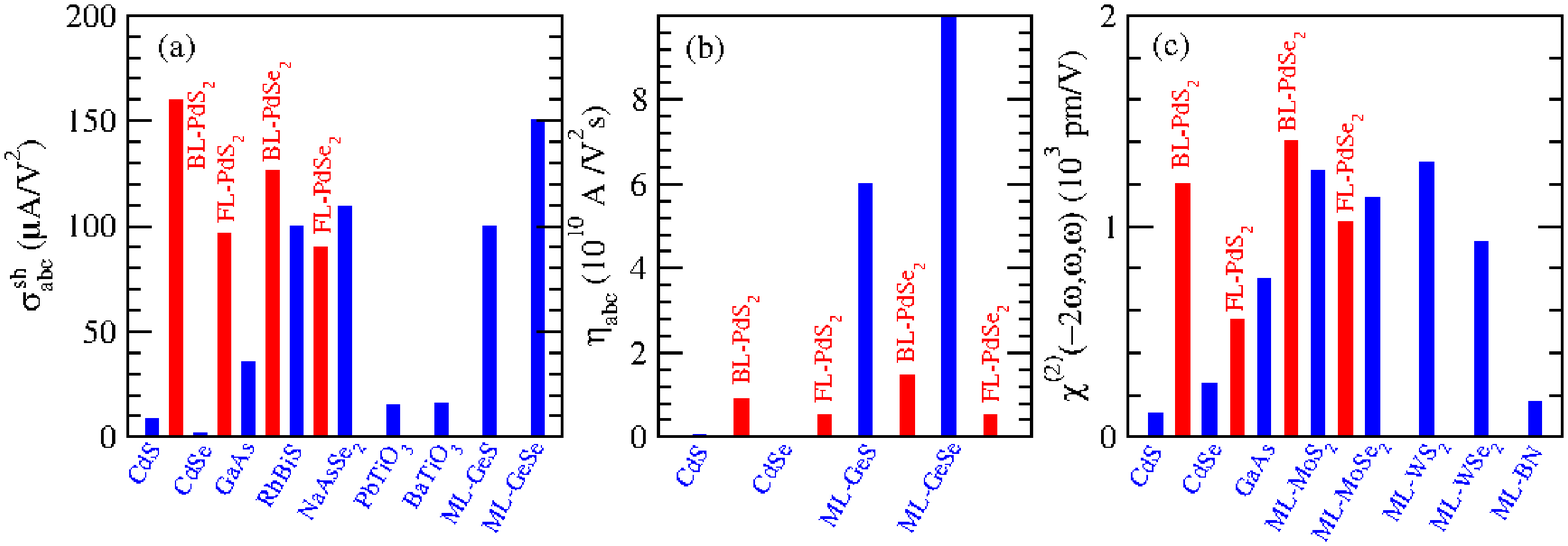}
\end{center}
\caption{Comparison of maximum nonlinear optical responses of the considered PdX$_2$ structures with other reported NLO materials
within the visible frequency range (i.e., up to 4.0 eV):
(a) Shift current conductivity~\cite{Rangel2017,Nastos2006,Young2012,Zhang2019b,Brehm2014,Laman2005,Ibanez-Azpiroz2018},
%Panday2019,Nastos2006,Laman2005,Zhang2019b,Young2012},
(b) injection current susceptibility~\cite{Panday2019,Nastos2006,Laman2005}
and (c) SHG susceptibility~\cite{Wang2015b,Huang1993,Bergfeld2003}.}
% \cite{Rangel2017,Young2012,Panday2019,Wang2019b,Laman2005,Song2018,Zhou2015,Sotome2019,Zhang2019b,Janisch2014,Li2013,Kumar2013,Brehm2014,Nastos2006}.}        
\label{comparison}
\end{figure}

Let us now compare the BPVE in the present structures with the well-known 
BPVE materials to access their application potentials in photovoltaic solar cells and opto-electronic devices.
The theoretically predicted shift current conductivity for the archetypal ferroelectrics PbTiO$_3$ 
and BaTiO$_3$~\cite{Young2012}, have a value within 10 $\mu$A/V$^2$ in the visible frequency range,
which is in agreement with the earlier experiments~\cite{Koch1976}.
% and 35 $\mu$A/V$^2$ above the photon energy 
%of 6 eV \cite{Young2012}. However, the spectra within the visible region have a maximum value 
%of 5 $\mu$A/V$^2$ for the above materials, which is much smaller compared to the present structures. 
%The experimental reported maximum value of the shift current response in the visible range 
%is 6 $\mu$A/V$^2$ for semiconductor SbSI \cite{Sotome2019}. 
These values are several times smaller than the present predictions for the 2D PdX$_2$ materials,
as shown in Fig. \ref{comparison}(a).
Recently, the shift current conductivity of some chiral materials was predicted 
to be rather large, being in the range of 20$\sim$80 $\mu$A/V$^2$ 
in the visible frequency range \cite{Zhang2019b}. 
%We have compared the present anomalous BPVE with traditional Si photocurrent, 
%it is 250 $\mu$A/V$^2$ which is two-times higher than the present shift current 
%of PdX$_2$ structures \cite{Rangel2017,Galagan2018}. 
Furthermore, ML group-IV monochalcogenides 
%have been attracting considerable attention in the field of BPVE because they 
were found to exhibit large shift current conductivity of about 100 $\mu$A/V$^2$ 
in the visible frequency range. Nevertheless, these values are smaller or at best comparable 
to the present predictions for the 2D PdX$_2$ structures (see Fig. \ref{comparison}).
%with ML monochalcogenides which shows a large value of about 100 $\mu$A/V$^2$ in the visible 
%spectrum \cite{Rangel2017}, which is lower compared to the present PdX$_2$ structures. 
%All the above comparison evidences that the considered 2D structures of PdX$_2$ have 
%a large shift current response in the visible range. 
Interestingly, the calculated injection current susceptibility of the present PdX$_2$ structures 
is two-orders larger than the experimental values of 1.5 $\times$ 10$^8$ A/V$^2s$ 
and 4 $\times $10$^8$ A/V$^2s$ for semiconductors CdSe and CdS~\cite{Laman2005}, respectively, 
which are in the same order of magnitude with the theoretical predictions reported in Ref. \cite{Nastos2010}.
% compared to the experimental 
%reported semiconductors of bulk Cd-based chalcogenides which are in the order of 
%1.5 $\times$ 10$^8$ A/V$^2s$ and 4 $\times $10$^8$ A/V$^2s$ for CdSe and CdS, respectively\cite{Laman2005}. 
Among the 2D materials, as Fig. \ref{comparison}(b) shows, the injection current susceptibility of ML group-IV 
chalcogenides was reported to be one-order larger compared to the present PdX$_2$ 
structures\cite{Panday2019,Wang2019b}. 

%Nonetheless, it is noteworthy to mentioned that the present injection current tensors are from the PdX$_2$ BLs and FLs structures. 
%%Overall, PdS$_2$ and PdSe$_2$ BLs and FLs structures will be a promising material for high efficient photovoltaic solar cell applications.
%The large shift and injection currents response contribution to the BPVE in PdX$_2$ structures 
%are attributed due to its low-dimensional, size reduction and strong covalent bonding, 
%which are known to enhance optical responses in the low-dimension systems \cite{Spanier2016,Alexe2011}. 
%The present study may find interesting applications in BPVE based photovoltaic materials and 
%we hope it will trigger experimental studies to search for promising candidates in these directions.

\begin{figure}[htb]
\begin{center}
\includegraphics[width=13cm]{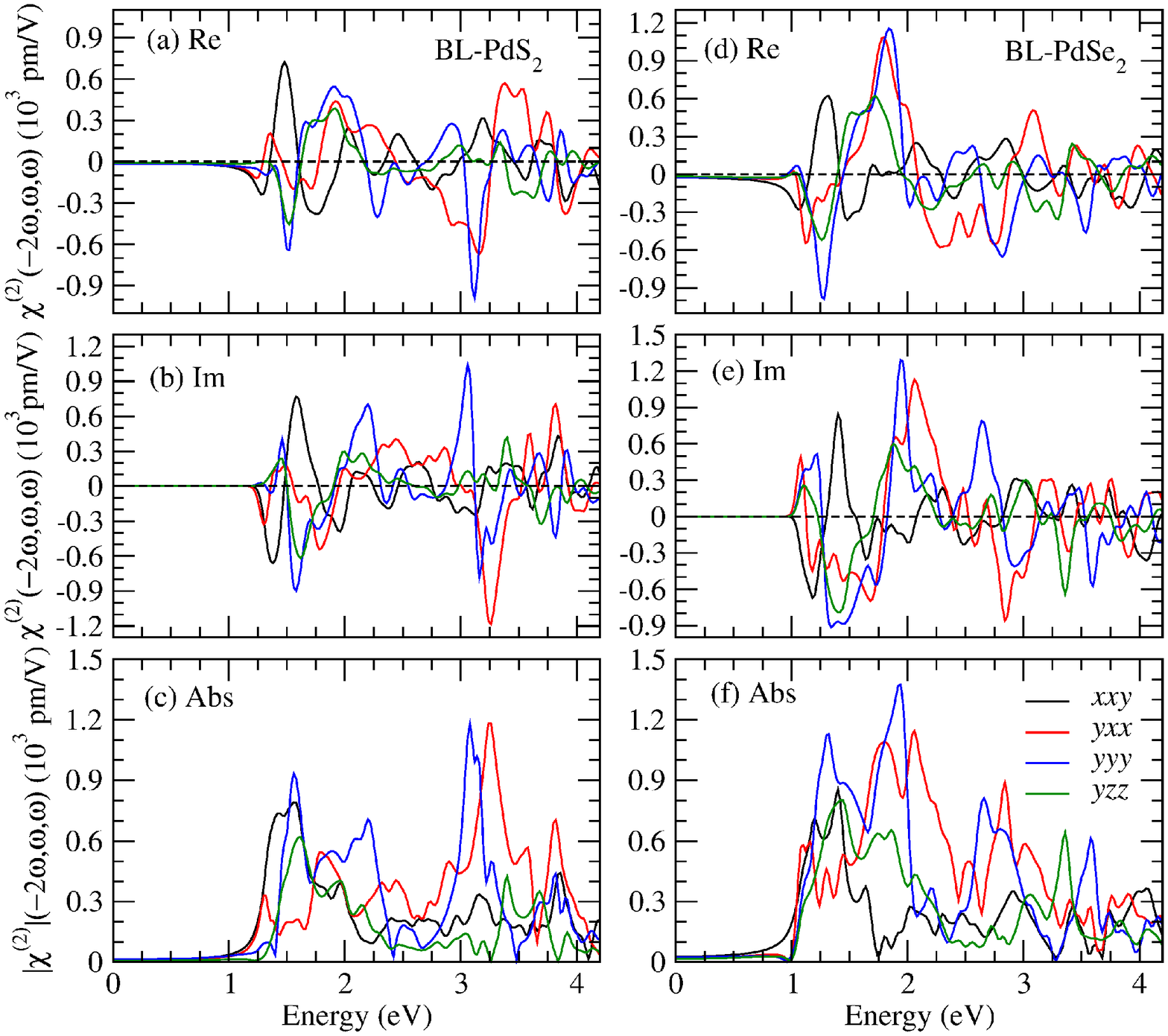}
\end{center}
\caption{Real (upper panels) and imaginary (middle panels) parts as well as modulus (lower panels) 
of the SHG susceptibility of BL PdS$_2$ and BL PdSe$_2$.}
\label{shg-re-im-BL}
\end{figure}

\begin{figure}[htb]
\begin{center}
\includegraphics[width=13cm]{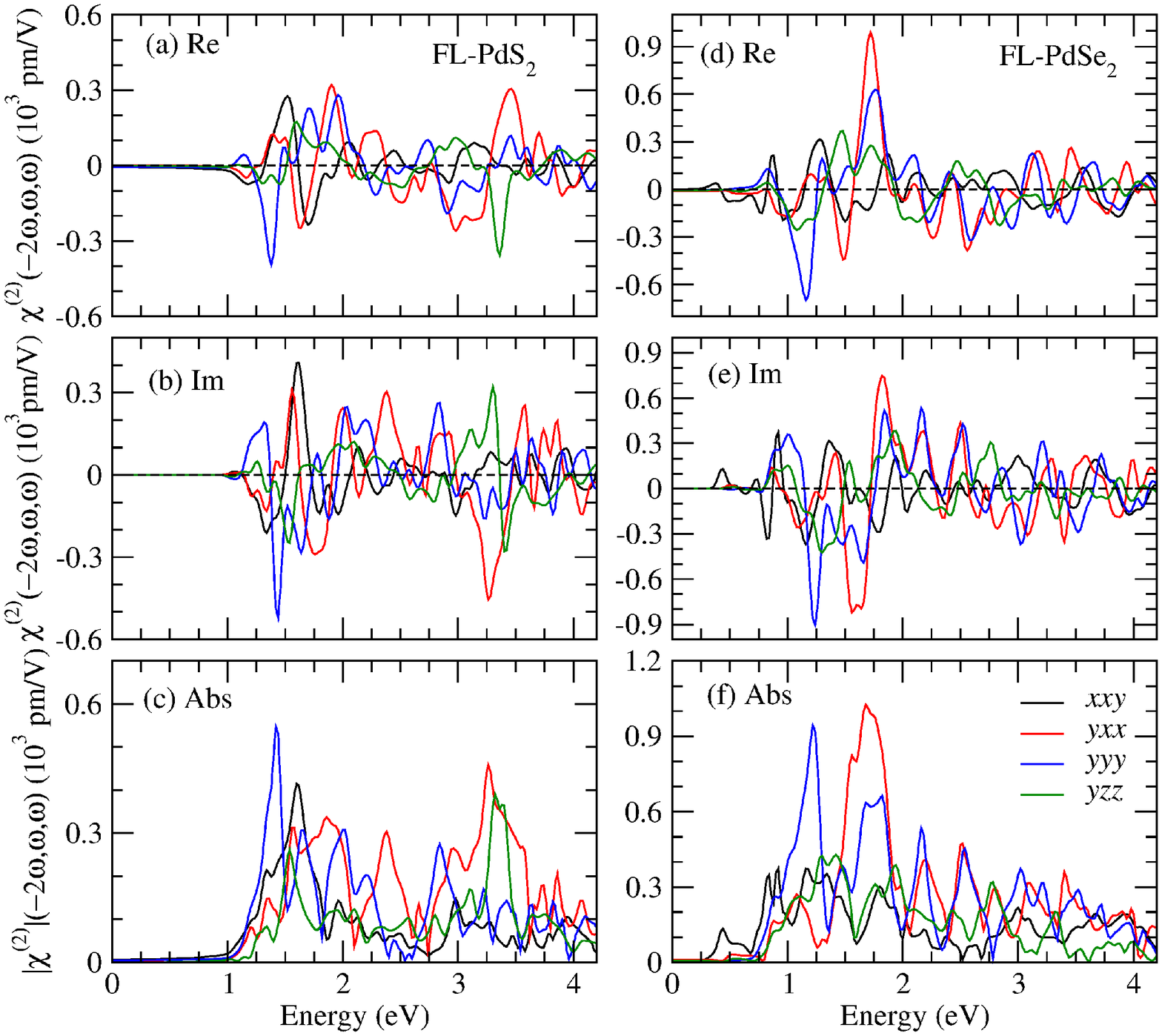}
\end{center}
\caption{Real (upper panels) and imaginary (middle panels) parts as well as modulus (lower panels)
of the SHG susceptibility of FL PdS$_2$ and FL PdSe$_2$.}
\label{shg-re-im-FL}
\end{figure}

\subsection{Second-harmonic generation}
For a noncentrosymmetric material, the nonzero elements of its SHG susceptibility tensor
are the same as that of its shift current conductivity tensor. 
Therefore, as for the shift current conductivity, the SHG susceptibility tensor of 
the BL and FL PdX$_2$ has five independent nonzero elements,  
i.e., $\chi^{(2)}_{xxy}=\chi^{(2)}_{xyx}$, $\chi^{(2)}_{yxx}$, 
$\chi^{(2)}_{yyy}$, $\chi^{(2)}_{yzz}$ and $\chi^{(2)}_{zyz}=\chi^{(2)}_{zzy}$ \cite{Shen2003,Boyd2003}. 
The real and imaginary parts as well as the modulus of these nonzero elements for BL PdX$_2$ and FL PdX$_2$ 
are presented in Fig.~\ref{shg-re-im-BL} and Fig.~\ref{shg-re-im-FL}, respectively.  
%The absolute values of these nonzero elements of BL PdX$_2$ 
%and FL PdX$_2$ are plotted in Fig.~\ref{BL-SHG} and                
%Fig.~\ref{FL-SHG}, respectively, along with the imaginary part of the corresponding dielectric function. 
Figures ~\ref{shg-re-im-BL} and Fig.~\ref{shg-re-im-FL} show that the imaginary (absorptive) parts of the SHG susceptibility  
for both BL and FL PdX$_2$ are zero for photon energy being smaller than half of the band gaps
but they increase rapidly above half of the band gaps.
Furthermore, below half of the band gaps, the real (dispersive) parts of the SHG susceptibility 
are small and remain almost constant. These nonzero dispersive parts of the SHG susceptibility 
below half of the band gaps give rise to the low-frequency LEO effect in the PdX$_2$ structures,
which will be discussed in the next section.
As for the imaginary parts, they increase rapidly above half of the band gaps. 

Among the nonzero SHG susceptibility elements, $\chi^{(2)}_{yxx}$
and $\chi^{(2)}_{yyy}$ generally exhibit larger magnitudes in the visible frequency
range for all the PdX$_2$ structures [see Figs. ~\ref{shg-re-im-BL}(c) and ~\ref{shg-re-im-BL}(f) as well
as ~\ref{shg-re-im-FL}(c) and ~\ref{shg-re-im-FL}(f)]. 
%The largest contribution to the $\chi^{(2)}$(-2$\omega,\omega,\omega$) are purely from 
%the dispersive part for the photon energy smaller than half of the band gap, because the absorptive 
%part of $\chi^{(2)}$ becomes nonzero only for photon energy larger than half of the band gap. 
%The dispersive part of $\chi^{(2)}$ below half of the band gap results in low-frequency 
%linear electro-optical coefficients, which we will discuss in the next section.
In BL PdS$_2$, both $\chi^{(2)}_{yyy}$ and $\chi^{(2)}_{yxx}$ spectra
show a pronounced peak with almost the same maximum value of $\sim$1.2$\times$10$^3$ pm/V
at slightly different photon energies of 3.1 and 3.3 eV, respectively [see Fig. ~\ref{shg-re-im-BL}(c)]. 
They come from the negative peak in the real and imaginary parts of the $\chi^{(2)}_{yyy}$ and $\chi^{(2)}_{yxx}$ spectra
at 3.1 and 3.3 eV, respectively [see Figs. ~\ref{shg-re-im-BL}(a) and ~\ref{shg-re-im-BL}(b)].
The $\chi^{(2)}_{yyy}$ spectrum of BL PdS$_2$ exhibits the second maximum of 0.9 $\times$10$^3$ pm/V 
at 1.6 eV [see Fig. ~\ref{shg-re-im-BL}(c)]. It originates from the negative peak near 1.6 eV
in both real and imaginary parts of the $\chi^{(2)}_{yyy}$ spectrum [see Figs. ~\ref{shg-re-im-BL}(a) and ~\ref{shg-re-im-BL}(b)]. 
Remarkably, the $\chi^{(2)}_{yyy}$ modulus spectrum of BL PdSe$_2$ 
has two gigantic peaks of heights of 1.1 and 1.4 $\times$10$^3$ pm/V at 1.4 and 1.9 eV, respectively
[see Figs. ~\ref{shg-re-im-BL}(f)] and they stem from the negative and positive peaks at 1.4 and 1.9 eV
in the real and imaginary parts of the spectrum [see Figs. ~\ref{shg-re-im-BL}(d) and ~\ref{shg-re-im-BL}(e)].
%1.4$\times$10$^3$ pm/V  is from $\chi^{(2)}_{yyy}$ in the order of 1.0 $\times$10$^3$ pm/V at the photon energy 
%of 1.6 eV (see Fig. 9(a)). 
%The higher value of the SHG spectra of the BL PdSe$_2$ structure is found at a photon energy 
%of 1.9 eV, which is 1.4$\times$10$^3$ pm/V for $\chi^{(2)}_{yyy}$. 
%The same spectra also shown another maximum value of 1.1$\times$10$^3$ pm/V at a photon energy 
%of 1.4 eV (see Fig. 9(c)). 
The $\chi^{(2)}_{yxx}$ spectrum of BL PdSe$_2$ has a broad twin peak of magnitude of ~1.1 $\times$10$^3$ pm/V 
centered at $\sim$1.9 eV [see Figs. ~\ref{shg-re-im-BL}(f)].
Nonetheless, below the band gap, the magnitudes of the $\chi^{(2)}_{xxy}$ and $\chi^{(2)}_{yzz}$ spectra
of the PdX$_2$ structures are generally comparable or even larger than that of the $\chi^{(2)}_{yyy}$ and $\chi^{(2)}_{yxx}$ spectra.
For example, both BL PdS$_2$ and BL PdSe$_2$ have a rather broad peak of $\sim$0.8$\times$10$^3$ pm/V 
at 1.5 and 1.3 eV, respectively [see Figs. ~\ref{shg-re-im-BL}(c) and ~\ref{shg-re-im-BL}(f)]. 
%is from $\chi^{(2)}_{yxx}$ has reached a substantial value 
%of 1.2$\times$10$^3$ pm/V at the same photon energy of 2.1 eV. 
All the $\chi^{(2)}_{abc}$ spectra from the FL PdX$_2$ structures
are generally smaller than the corresponding spectra of the BL PdX$_2$ structures.
Nonetheless, the magnitudes of the $\chi^{(2)}_{yyy}$ and $\chi^{(2)}_{yxx}$ spectra of FL PdSe$_2$
do peak at 1.2 and 1.7 eV with the large maximum values of 0.94$\times$10$^3$ and 1.2$\times$10$^3$ pm/V,
respectively [see Fig. ~\ref{shg-re-im-FL}(f)]. 

\begin{figure}[htb]
\begin{center}
\includegraphics[width=13cm]{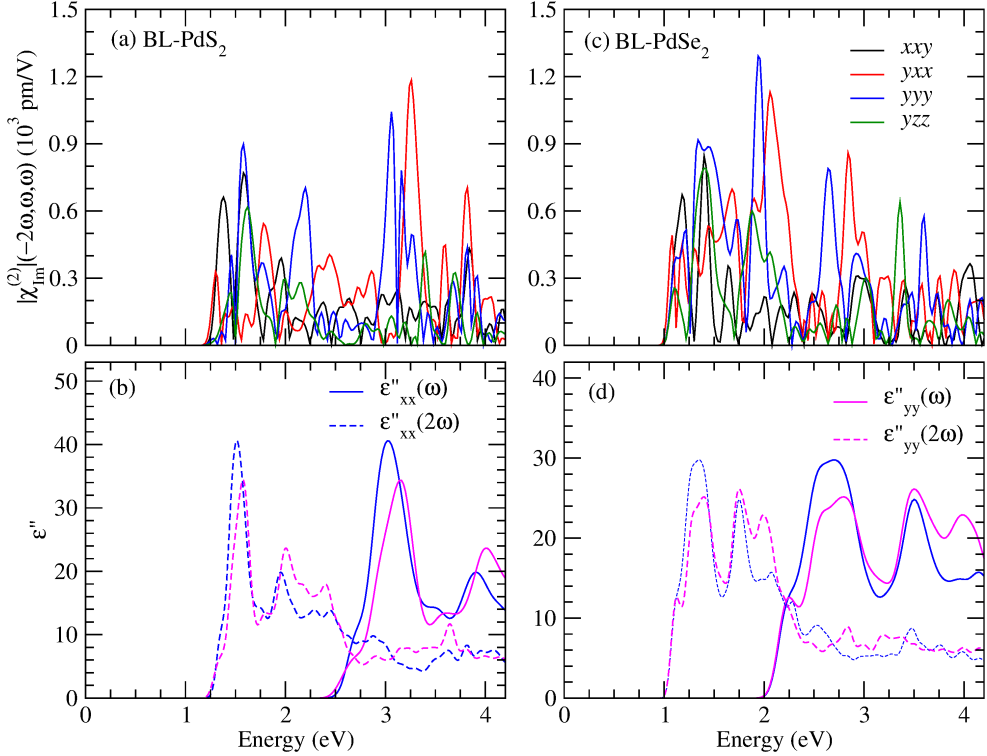}
\caption{(a)[(c)] Absolute SHG susceptibility and (b)[(d)] imaginary dielectric constant of BL PdS$_2$ [PdSe$_2$].}
\end{center}
\label{BL-SHG}
\end{figure}

As shown before (see, e.g., ~\cite{Wang2015b} and references therein), the prominent features in the SHG susceptibility 
are generally caused by either single ($\omega$) and double (2$\omega$) photon resonances or both. 
This can also be seen from Eqs. (7) and (8).
Thus, to help understand the origins of the prominent features in the calculated SHG 
spectra, we plot the absolute values of the imaginary parts of the nonzero SHG elements of BL PdX$_2$ and FL PdX$_2$
in Fig. 10 and Fig. 11, respectively, 
along with the absorptive parts of the corresponding dielectric functions $\varepsilon''(\omega)$ and $\varepsilon''(2\omega)$.
%As per the definition of SHG spectra, it consist of both single ($\omega$) and 
%double (2$\omega$) photon resonances. 
%It is very important to analyze the prominent features of single and double photons 
%in the calculated SHG spectra. 
%To this end, we have presented single and two-photon absorptive parts of the dielectric 
%function $\varepsilon''(\omega)$ and $\varepsilon''(2\omega)$ together in the Fig. 9(b), (d) and Fig. 10(b), (d). 
Figures 10(a) and 10(b) [11(a) and 11(b)] show that the prominent features in the $|\chi''^{(2)}|$ spectra 
of BL (FL) PdS$_2$ in the energy range 
of 1.2-2.5 eV (1.1-2.2 eV) below the band gap look similar to 
% from the absorption edge of $\varepsilon''(2\omega)$ to the absorption edge of $\varepsilon''(\omega)$ 
the features in the $\varepsilon''(2\omega)$ spectra, 
%i.e. from photon energy of 1.2-2.5 eV (1.1-2.2 eV), 
indicating that they are due to double-photon resonances. 
Similarly the pronounced features in the $|\chi''^{(2)}_{yyy}|$ and $|\chi''^{(2)}_{yxx}|$ spectra above the band gap of 
2.5 eV (2.2 eV), have a shape similar to that in the $\varepsilon''(\omega)$, 
suggesting that they are caused primarily by single-photon resonances. 
On the other hand, the $|\chi''^{(2)}_{yyy}|$ and $|\chi''^{(2)}_{yxx}|$ spectra above the band gap of                  
2.5 eV (2.2 eV) have a much reduced amplitude and also are rather oscillatory,
being rather similar to the $\varepsilon''(2\omega)$ spectra in this regime.
This indicates that they stem mainly from the double (2$\omega$) photon resonances. 

%In the case of BL (FL) PdSe$_2$ structures the peaks in the $|\chi"^{(2)}|$ spectra in the energy range 
%of 1.0-2.0 eV (0.75-1.7 eV) from the double-photon resonance and the peaks above the absorption edge 
%of $\varepsilon''(\omega)$, i.e. above 2.0 eV (1.7 eV), are due to either $\varepsilon''(2\omega)$ 
%or $\varepsilon''(\omega)$ or both. Because of the contributions from both one and two-photon resonances, 
%the spectra oscillate rapidly in this region and diminish gradually at higher photon energies. 
%Almost all the SHG susceptibility elements have the same oscillatory behavior. 
%The peak maximum value of $\chi"^{(2)}_{yyy}$ and $\chi"^{(2)}_{yxx}$ of BL (at 3.1 eV) PdS$_2$ structures 
%are due to the contribution from both single and two-photon resonances. 
%In the case of BL PdSe$_2$ the maximum value of $\chi"^{(2)}_{yyy}$ at the photon energy of 1.9 eV 
%is due to the contribution from double-photon resonances. 
%For FL PdS$_2$ structure the peak maximum at 1.4 eV from $\chi"^{(2)}_{yyy}$ is due to the 
%double-photon resonances.
%In the case of FL PdSe$_2$ the maximum value of $\chi"^{(2)}_{yxx}$ at the photon energy of 1.7 eV 
%is due to the contribution from both single and two-photon resonances.

\begin{figure}[htb]
\begin{center}
\includegraphics[width=13cm]{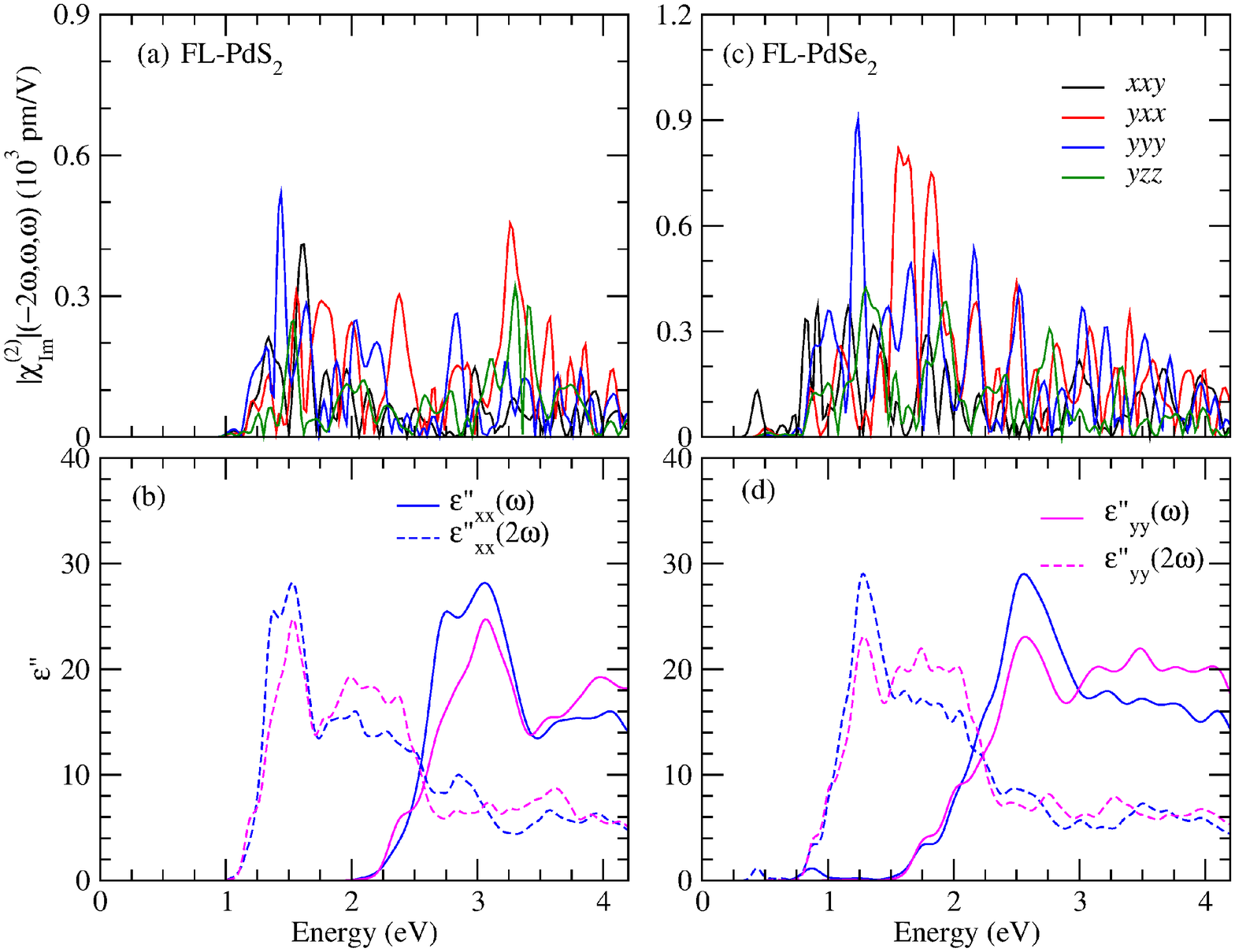}
\caption{(a)[(c)] Absolute SHG susceptibility and (b)[(d)] imaginary dielectric constant of FL PdS$_2$ [PdS$e_2$].}
\end{center}
\label{FL-SHG}
\end{figure}

\begin{table*}[htbp]
\begin{center}
\caption{Calculated static dielectric constants ($\varepsilon_x$, $\varepsilon_y$ and $\varepsilon_z$), 
second-order optical susceptibility $\chi^{(2)}$(0) in pm/V, and linear electro-optical 
coefficients $r_{abc}$ in pm/V of BL and FL PdX$_2$ structures.}
\begin{tabular}{ccccccccc}
%\br %\hline
\hline \hline
      			&$\varepsilon_x$	&$\varepsilon_y$  &$\varepsilon_z$ &	$xxy$	&$yxx$	&$yyy$	&$yzz$	\\
\hline
BL-PdS$_2$		&12.95	       	&12.73	&     5.97 \\
$\chi^{(2)}$(0)(pm/V)	&	&	        	&		&-11	&-8.80	&-13	&-7.04	\\

$r_{abc}$(pm/V)	&	&			&		&0.13	&0.11	&0.16	&0.09	\\
\hline
BL-PdSe$_2$		&13.84	      		&13.90	&     6.68 \\
$\chi^{(2)}$(0)(pm/V)	&	&	&				&-28	&-22	&-24	&-18	\\
$r_{abc}$(pm/V)	&	&		&			&0.29	&0.23	&0.25	&0.19	\\
\hline
FL-PdS$_2$		&12.09       		&11.92	&     5.40 \\
$\chi^{(2)}$(0)(pm/V)	&	&	&		        	&-6.75	&1.39	&-5.41	&-0.42	\\

$r_{abc}$(pm/V)	&	&		&			&0.09	&-0.02	&0.08	&0.01	\\
\hline
FL-PdSe$_2$		&14.84	      		&14.64	&     8.46 \\
$\chi^{(2)}$(0)(pm/V)	&	&	&				&-0.96	&-12	&-4.90	&-5.78	\\
$r_{abc}$(pm/V)	&	&		&			&0.01	&0.11	&0.05	&0.05	\\
\hline \hline
\end{tabular} \\
\end{center}
\label{table}
\end{table*}

We now compare the calculated SHG susceptibility of BL and FL PdX$_2$ with that reported for other NLO materials.
% to see how promising they are for the NLO applications. 
%Let us first compare the calculated SHG susceptibility of the present PdX$_2$ structures with that of 
Monolayer group 6B TMDC semiconductors such as MoS$_2$,
are considered to be the most promising 2D NLO materials because of their
direct band gaps and large SHG susceptibility (see ~\cite{Wang2015b} and references therein).
%The experimental reported $\chi^{(2)}$ of mechanically exfoliated ML MoS$_2$ is as large
%as $\sim$10$^5$ pm/V at 810 nm wavelength for chemical vapor deposition of the same material,
%the $\chi^{(2)}$ value was reported as 5$\times$10$^3$ pm/V \cite{Kumar2013}.
%In another experimental work the value of $\chi^{(2)}$ is about 320 pm/V\cite{Li2013}.
%In the case of few-layers PdSe$_2$ structures the $\chi^{(2)}$ shows a maximum value
%1.4$\times$10$^3$ pm/V (1.0$\times$10$^3$ pm/V) at a photon energy of 1.9 eV (1.7 eV)
%for BL (FL) structures and PdS$_2$ BL shown a maximum value of 1.2$\times$10$^3$ pm/V,
%which can be large compared to ML MoS$_2$ \cite{Li2013,Malard2013},
%but comparable to 4.5$\times$10$^3$ pm/V of single-layer WS$_2$ \cite{Janisch2014}.
Remarkably, Fig. 7(c) shows that the maximum values of the SHG susceptibility of BL PdX$_2$ 
are comparable or even larger than that of ML group 6B TMDCs~\cite{Wang2015b}.
Similarly, FL PdX$_2$ generally have the SHG susceptibility that are comparable 
or even larger than that of TL group 6B TMDCs~\cite{Wang2015b}.
%Note that only odd-number few-layer group 6B TMDCs have nonzero SHG coefficients.
%As mentioned above, there are few reported even-number few-layer TMDCs 
%which exhibit NLO responses~\cite{Song2018,Zhou2015,Hu2017b}. 
Interestingly, ReS$_2$ is another rare TMDC that exhibits second-order NLO responses 
only when the number of layers is even.
% also has non-zero SHG susceptibility in its even layers and zero SHG susceptibility in the odd layers.
The measured SHG susceptibility of BL ReS$_2$ is large, being about 900 pm/V at 0.8 eV~\cite{Song2018}. 
Nevertheless, this is smaller than that of BL PdS$_2$ and BL PdSe$_2$. 

\subsection{Linear electro-optical coefficient}
%The linear electro-optic effect is the change in the optical dielectric constant due to an applied electric field.
%%The LEO effect has been widely exploited in the integrated optical devices for optical communications and
%light modulation. 
%Thus, based on the obtained SHG susceptibility at low-frequency limit
%and static dielectric constant, 
Here we estimate the LEO coefficients of BL and FL PdX$_2$ structures, 
based on the obtained SHG susceptibility at low-frequency limit and static dielectric constant. 
%The LEO effect is considered as a special effect of second-order NLO properties and 
%has been used extensively in the integrated optical devices for optical communications and modulating light. 
%The origin of the LEO effect is the change in the optical dielectric constants due to applied electric field. 
Note that the LEO coefficients we present here represent only the electronic contribution.
%The second-ordered nonlinear spectrum based LEO coefficients are considered as 
%the electronic contribution to the LEO coefficient. 
There are other contributions to the LEO coefficient such as ionic and piezoelectric contributions, 
which are beyond of the scope of the present work~\cite{Veithen2005}.
The calculated LEO coefficients $r_{abc}(0)$ at zero frequency along with 
the static dielectric constants and SHG susceptibilities $\chi^{(2)}_{xyz}(0,0,0)$ are listed in Table 2. 
%The obtained BL and FL PdX$_2$ structures static SHG susceptibilities $\chi^{(2)}_{xyz}(0,0,0)$ 
%exhibit strong anisotropy in their spectrum. 
It is clear from Table 2 that BL PdS$_2$ and BL PdSe$_2$ have much larger LEO values
than FL PdS$_2$ and FL PdSe$_2$. BL PdS$_2$ and BL PdSe$_2$ also exhibit a rather strong anisotropy in the LEO effect. 
Semiconductor GaAs was reported to have a LEO coefficient of $r_{xyz}(0)=-1.5$ pm/V~\cite{Johnston1969}.
Recent calculations~\cite{Wang2015b} predicted that the magnitudes of the LEO coefficients of monolayer 
group 6B TMDC semiconductors are about 1.5 pm/V, being close to that of GaAs.
Table 2 indicates that the LEO coefficients for BL PdS$_2$ and BL PdSe$_2$ are many times smaller than
that of GaAs~\cite{Johnston1969} and also monolayer group 6B TMDC semiconductors~\cite{Wang2015b}.
Nevertheless, they are in the same order of magnitude as that of trilayer group 6B TMDC semiconductors~\cite{Wang2015b}.
%They are also comparable to that of single BN sheet~\cite{Guo2005}.
%the calculated low-frequency static SHG susceptibilities 
%of $\chi^{(2)}_{xyz}(0,0,0)$ for both BL PdS$_2$ and PdSe$_2$ structures are significant.  
%However, the low-photon energy LEO coefficients of all the structures of PdX$_2$ 
%are found to be lower in value compared to other 2D TMDC materials.  
%In particular, the LEO coefficients of both PdS$_2$ adn PdSe$_2$ BL structure 
%are almost three times lower than that of 2D single-layer Mo-based TMDCs and comparable 
%to the tri-layer structures of the same materials \cite{Wang2015b}. 
%Though the LEO coefficients of the present structures are very low compared to the other materials, 
%we stress that the LEO coefficients are due to the SHG susceptibility of the even layer systems, 
%which are unique for these structures.
%The anisotropic nature is also present in the LEO coefficients. 
%As of now there are no reports available for any even layer materials to be compared 
%with the present structures on LEO coefficients.
%The calculated LEO coefficients of PdX$_2$ structures may find excellent applications 
%in linear electro-optical modulators.

\section{Discussion and conclusions}
\begin{figure}[htb]
\begin{center}
\includegraphics[width=7.5cm]{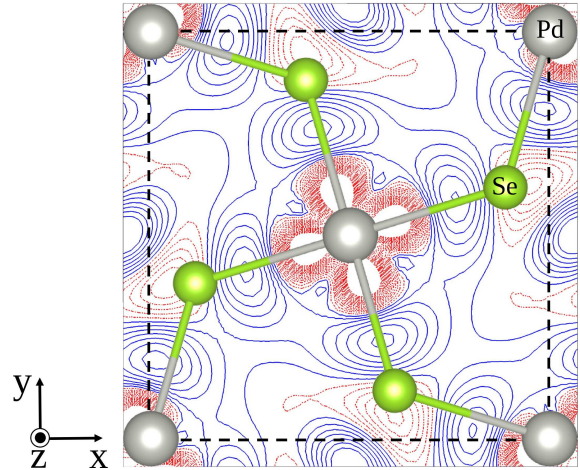}
\end{center}
\caption{Contour plot of the deformation charge density distribution of BL PdSe$_2$.
The contour interval is 0.002 e/\AA$^3$. The electron accumulation is depicted by positive contours
(blue solid lines), while the electron depletion is represented by negative contours (red dashed lines).
The black dashed line indicates the 2D unit cell [see Fig. 1(c)].}
\label{dcd}
\end{figure}

It was recently pointed out~\cite{Cheng2019} that the size of the band gap of a semiconductor 
is a principal factor that determines the strength of its second-order NLO responses. 
To understand other origins of the large NLO responses in the 2D PdX$_2$ structures compared
with other NLO materials of similar band gaps, 
we calculate the deformation charge density, which is defined as the difference 
between the valence charge density and the superposition of the free atomic charge densities. 
The calculated deformation charge density distributions for the four 2D PdX$_2$ structures
look very similar. Thus, here we focus only on the deformation charge density of BL PdSe$_2$, 
which is displayed in Fig. \ref{dcd}.
%The deformation charge density of the BL PdSe$_2$ structure is presented in Fig. \ref{dcd}.
Clearly, there is a significant buildup of the electron charge in the vicinity of the Pd-Se bond center
by depleting the charge around the Pd atoms along the bond directions. This is caused
by the strong directional covalent bonding in BL PdSe$_2$, and can lead to an enhanced
optical responses due to large spatial overlap between the wavefunctions of the initial and final states,
and high anisotropy which would result in large NLO response values \cite{Cheng2019,Song2009,Ingers1988}.
%The semiconducting nature along with strong covalency makes few-layer PdX$_2$ structures
%possess a large value of dc photoconductivity and SHG susceptibility.

In conclusion, we have systematically studied the second-order NLO responses of 
BL and FL PdX$_2$ structures using first-principles DFT calculations. 
First of all, we predict that the BPVE in the considered 2D PdX$_2$ structures is generally
strong, with a large shift current conductivity of up to 130 $\mu$A/V$^2$
and injection current susceptibility of up to 100$\times$10$^8$ A/V$^2$s in the visible frequency range.
Secondly, we find that all PdX$_2$ structures possess large SHG susceptibility $\chi^{(2)}$ in the visible frequency spectrum
and BL PdSe$_2$ has the largest $\chi^{(2)}$ value of 1.4$\times$10$^3$ pm/V at 1.9 eV.
Thirdly, we find that the LEO coefficients of BL PdS$_2$ and BL PdSe$_2$ are significant.
%and comparable to that of trilayer group 6B TMDC semiconductors~\cite{Wang2015b}.
%The BPVE (also known as photogalvanic effect) refers to the generation
%of dc photocurrents in noncentrosymmetric materials.~\cite{Sturman1992}
%In a nonmagnetic semiconductor, there are two main contributions to the BPVE,
%namely, the circular injection current and linear shift current~\cite{Sturman1992,Sipe2000,Ahn2020,Gudelli2020}
%Materials having large BPVE are crucial for applications in photovoltaic solar cells and high sensitive photodetectors.
%Here we predict that the BPVE in the considered few-layer PdX$_2$ structures is generally
%strong, with a large shift current conductivity of up to 130 $\mu$A/V$^2$
%and injection current susceptibility of up to 100$\times$10$^8$ A/V$^2$s in the visible frequency range.
These superior NLO responses of the BL and FL PdX$_2$ structures
will make them valuable for technological applications in NLO and electo-optic
devices such as light signal modulators, frequency converters, electro-optical switches,
photovoltaics and photodetector applications.
Finally, the strong NLO responses of BL and FL structures of PdX$_2$ are attributed to 
strong intralayer directional covalent bonding and also 2D quantum confinement.

\textbf{Acknowledgement:}
The authors acknowledge the support by the Ministry of Science and Technology and 
National Center for Theoretical Sciences, Taiwan. The authors also thanks National Center 
for High-performance Computing, Taiwan for the computing time. 

%\newpage
\subsection*{References}
%\bibliography{2DM-PdSe2-BL-FL}

{}

\newpage

\begin{center}
{\bf --Supplementary information--}
\end{center}

\title{Large bulk photovoltaic effect and second-harmonic generation in few-layer 
pentagonal semiconductors PdS$_2$ and PdSe$_2$}
\author{Vijay Kumar Gudelli$^1$ and Guang-Yu Guo$^{2,1}$}
\address{$^1$ Physics Division, National Center for Theoretical Sciences, Taipei 10617, Taiwan}
\address{$^2$ Department of Physics and Center for Theoretical Physics, National Taiwan University, Taipei 10617, Taiwan}
\ead{gyguo@phys.ntu.edu.tw}

%\maketitle

\begin{figure*}[htb]
\begin{center}
\includegraphics[width=13cm]{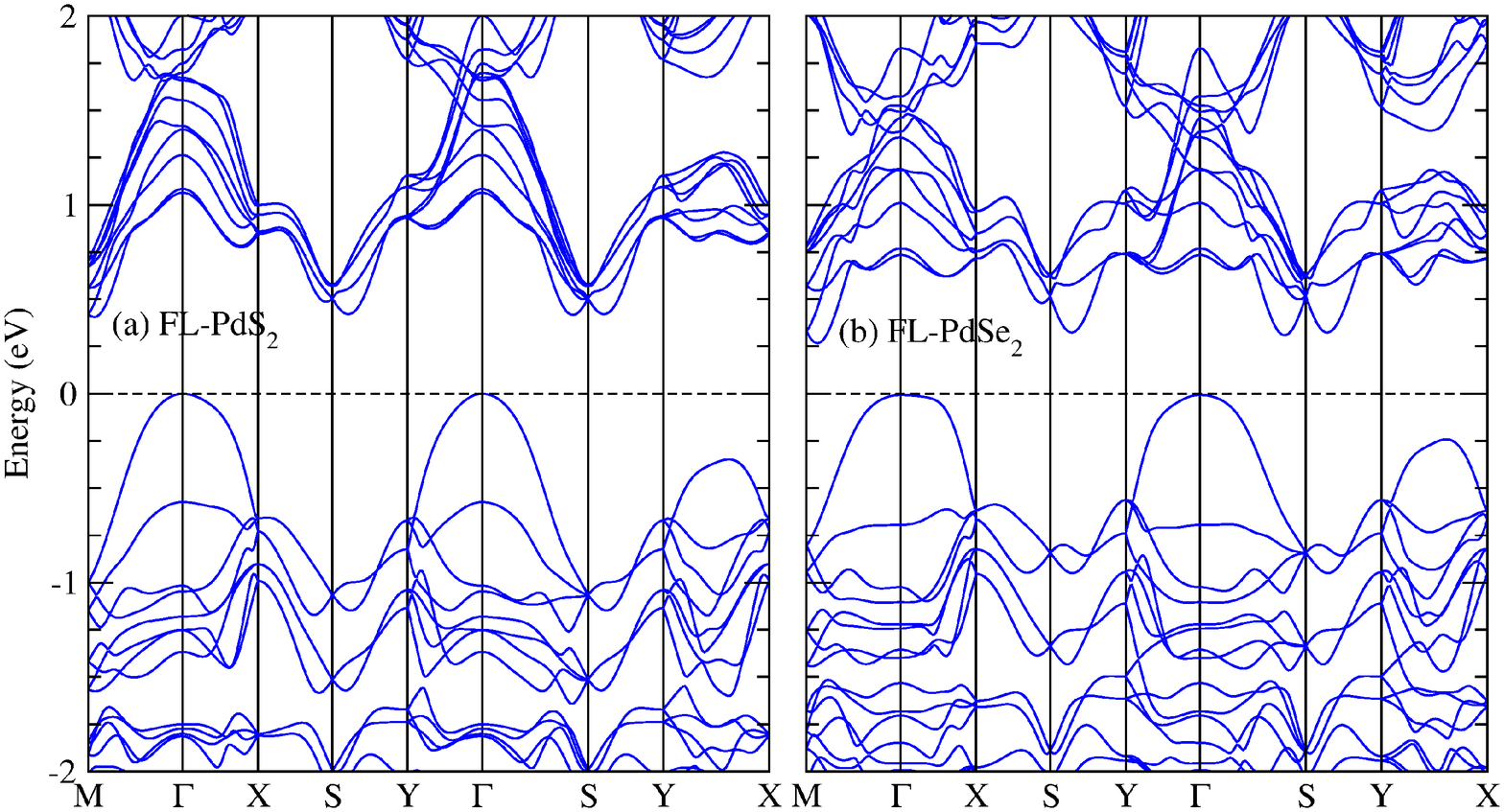}
\end{center}
\captionsetup{labelformat=empty}
\caption{Fig. S1: GGA-functional band structures of (a) FL PdS$_2$ and (b) FL PdSe$_2$. The horizontal black dashed line denotes the top of valance bands.}
%\label{FL-band}
\end{figure*}

\begin{figure*}[htb]
\begin{center}
\includegraphics[width=13cm]{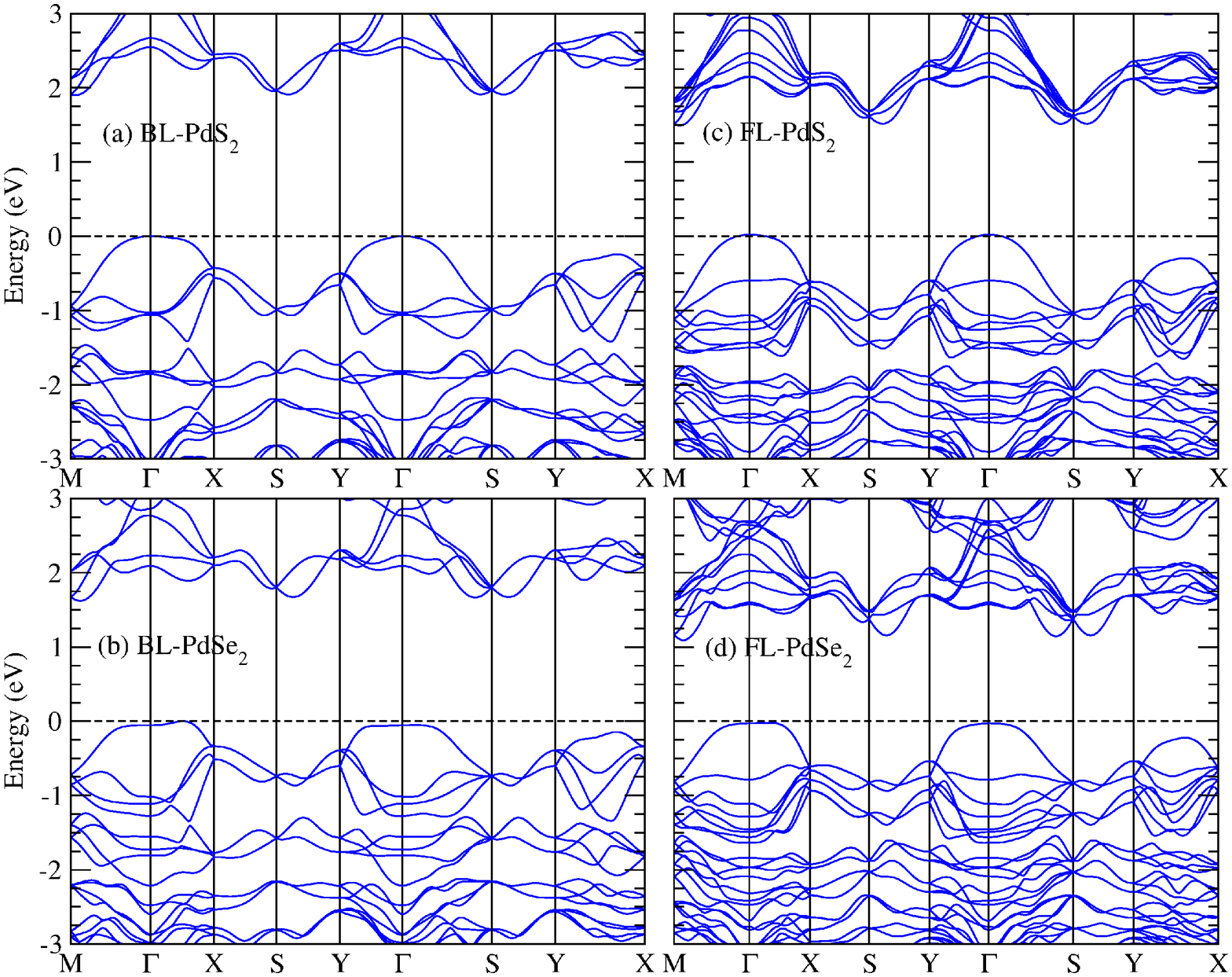}
\end{center}
\captionsetup{labelformat=empty}
\caption{Fig. S2: HSE-functional band structures of (a) BL and (c) FL PdS$_2$ as well as (b) BL and (d) FL PdSe$_2$. The horizontal black dashed line denotes the top of valance bands.}
%\label{FL-dos}
\end{figure*}

\begin{figure*}[htb]
\begin{center}
\includegraphics[width=13cm]{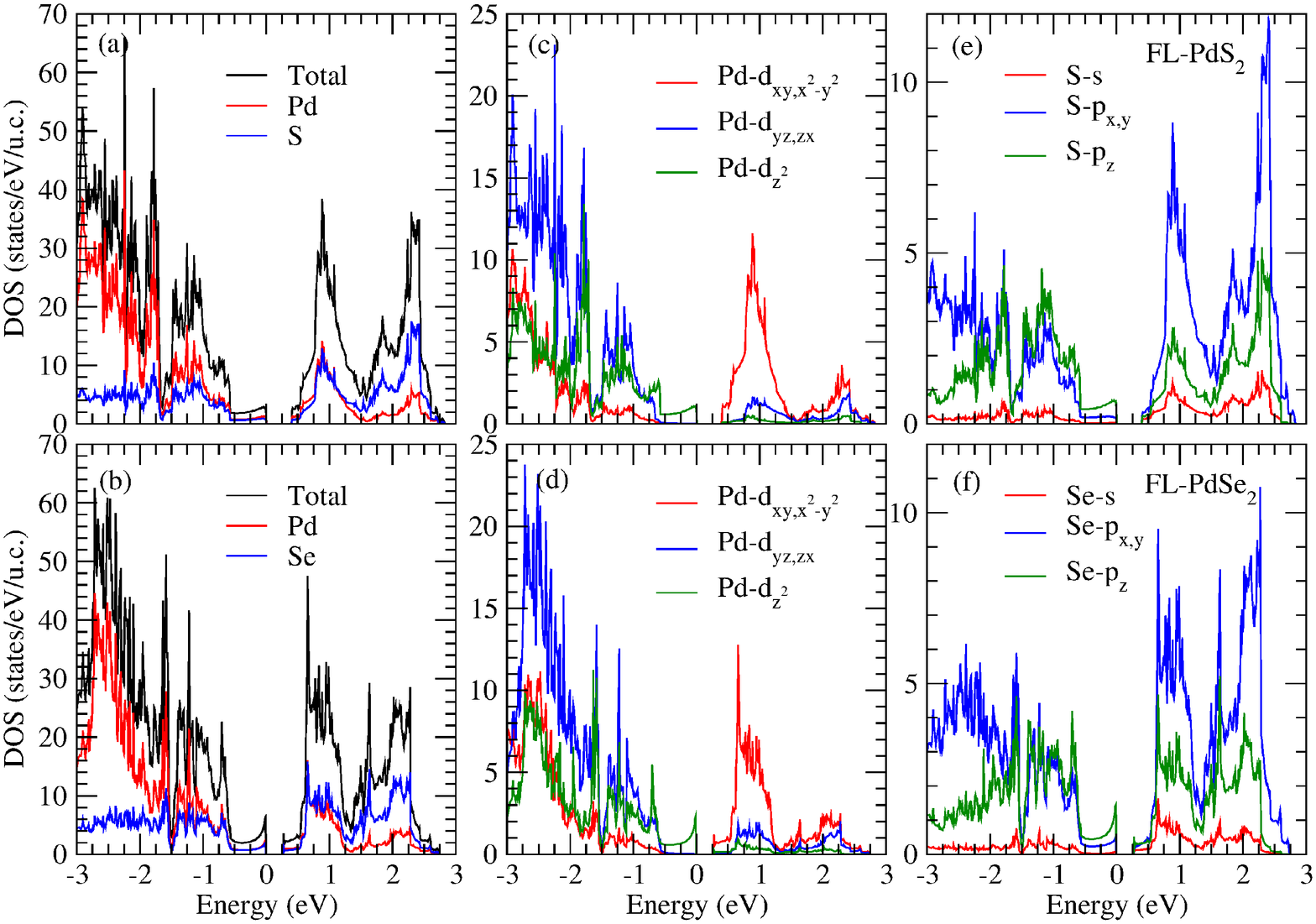}
\end{center}
\captionsetup{labelformat=empty}
\caption{Fig. S3: Total- and orbital-projected densities of states (DOS) for FL PdS$_2$ (upper panels) and FL PdSe$_2$ (lower panels) obtained using the GGA-functional.}
\end{figure*}

\end{document}